\documentclass[]{aastex631}
\newcommand{\fwd}{{\it forward}}
\newcommand{\bwd}{{\it backward}}

\usepackage{multirow}

\defcitealias{old_moon_model}{KP87}
\defcitealias{hafez_moon_model}{H14}

\begin{document}

\title{KISS: instrument description and performance}

\correspondingauthor{J.~F. Mac\'ias-P\'erez \& M. Fern\'{a}ndez-Torreiro}
\email{macias@lpsc.in2p3.fr; mateo@fernandeztorreiro.com}

\author[0000-0002-5385-2763]{J.~F. Mac\'ias-P\'erez}
\affiliation{Laboratoire de Physique Subatomique et de Cosmologie, Universit\'e Grenoble Alpes, CNRS/IN2P3, 53, avenue des Martyrs, Grenoble, France}
\author[0000-0002-6805-9100]{M. Fern\'{a}ndez-Torreiro}
\affiliation{Laboratoire de Physique Subatomique et de Cosmologie, Universit\'e Grenoble Alpes, CNRS/IN2P3, 53, avenue des Martyrs, Grenoble, France}
\affiliation{Instituto de Astrofísica de Canarias, C/Vía Láctea s/n, E-38205 La Laguna, Tenerife, Spain}
\affiliation{Departamento de Astrof\'{\i}sica, Universidad de La Laguna, 
    E-38206 La Laguna, Tenerife, Spain}
\author{A. Catalano}
\affiliation{Laboratoire de Physique Subatomique et de Cosmologie, Universit\'e Grenoble Alpes, CNRS/IN2P3, 53, avenue des Martyrs, Grenoble, France}
\author{A. Fasano}
\affiliation{Institut N\'eel, CNRS and Universit\'e Grenoble Alpes, France}
\affiliation{Instituto de Astrofísica de Canarias, C/Vía Láctea s/n, E-38205 La Laguna, Tenerife, Spain}
\affiliation{Departamento de Astrof\'{\i}sica, Universidad de La Laguna, 
    E-38206 La Laguna, Tenerife, Spain}
 \author{M. Aguiar}
\affiliation{Instituto de Astrofísica de Canarias, C/Vía Láctea s/n, E-38205 La Laguna, Tenerife, Spain}
\affiliation{Departamento de Astrof\'{\i}sica, Universidad de La Laguna, 
    E-38206 La Laguna, Tenerife, Spain}
	 \author{A. Beelen} 
\affiliation{Aix Marseille Universit\'e, CNRS, LAM (Laboratoire d'Astrophysique de Marseille) UMR 7326, 13388, Marseille, France}
	 \author{A. Benoit} 
\affiliation{Institut N\'eel, CNRS and Universit\'e Grenoble Alpes, France}
\author{A. Bideaud}
\affiliation{Institut N\'eel, CNRS and Universit\'e Grenoble Alpes, France}
\author{J. Bounmy}
\affiliation{Laboratoire de Physique Subatomique et de Cosmologie, Universit\'e Grenoble Alpes, CNRS/IN2P3, 53, avenue des Martyrs, Grenoble, France}
\author{O. Bourrion}
\affiliation{Laboratoire de Physique Subatomique et de Cosmologie, Universit\'e Grenoble Alpes, CNRS/IN2P3, 53, avenue des Martyrs, Grenoble, France}
\author{M. Calvo}
\affiliation{Institut N\'eel, CNRS and Universit\'e Grenoble Alpes, France}
\author{J.~A. Castro-Almaz\'an}
\affiliation{Instituto de Astrofísica de Canarias, C/Vía Láctea s/n, E-38205 La Laguna, Tenerife, Spain}
\affiliation{Departamento de Astrof\'{\i}sica, Universidad de La Laguna, 
    E-38206 La Laguna, Tenerife, Spain}
	 \author{P. de Bernardis}
\affiliation{Dipartimento di Fisica, Sapienza Universit\`a di Roma, Piazzale Aldo Moro 5, I-00185 Roma, Italy}
\author{M. De Petris}
\affiliation{Dipartimento di Fisica, Sapienza Universit\`a di Roma, Piazzale Aldo Moro 5, I-00185 Roma, Italy}
	\author{A. P. de Taoro}
\affiliation{Instituto de Astrofísica de Canarias, C/Vía Láctea s/n, E-38205 La Laguna, Tenerife, Spain}
\affiliation{Departamento de Astrof\'{\i}sica, Universidad de La Laguna, 
    E-38206 La Laguna, Tenerife, Spain}
	\author{G. Garde}
\affiliation{Institut N\'eel, CNRS and Universit\'e Grenoble Alpes, France}
    \author{R.~T. G\'{e}nova-Santos}
\affiliation{Instituto de Astrofísica de Canarias, C/Vía Láctea s/n, E-38205 La Laguna, Tenerife, Spain}
\affiliation{Departamento de Astrof\'{\i}sica, Universidad de La Laguna, 
    E-38206 La Laguna, Tenerife, Spain}
	\author{A. Gomez}
\affiliation{Centro de Astrobiolog\'{\i}a (CSIC-INTA), Torrejón de Ardoz, 28850 Madrid, Spain}
	\author{M. F. G\'{o}mez-Renasco} 
\affiliation{Instituto de Astrofísica de Canarias, C/Vía Láctea s/n, E-38205 La Laguna, Tenerife, Spain}
\affiliation{Departamento de Astrof\'{\i}sica, Universidad de La Laguna, 
    E-38206 La Laguna, Tenerife, Spain}
	\author{J. Goupy}
\affiliation{Institut N\'eel, CNRS and Universit\'e Grenoble Alpes, France}
	\author{C. Hoarau}
\affiliation{Laboratoire de Physique Subatomique et de Cosmologie, Universit\'e Grenoble Alpes, CNRS/IN2P3, 53, avenue des Martyrs, Grenoble, France}
\author{R. Hoyland}
\affiliation{Instituto de Astrofísica de Canarias, C/Vía Láctea s/n, E-38205 La Laguna, Tenerife, Spain}
\affiliation{Departamento de Astrof\'{\i}sica, Universidad de La Laguna, 
    E-38206 La Laguna, Tenerife, Spain}
	\author{G. Lagache}
\affiliation{Aix Marseille Universit\'e, CNRS, LAM (Laboratoire d'Astrophysique de Marseille) UMR 7326, 13388, Marseille, France}
	\author{J. Marpaud}
\affiliation{Laboratoire de Physique Subatomique et de Cosmologie, Universit\'e Grenoble Alpes, CNRS/IN2P3, 53, avenue des Martyrs, Grenoble, France}
\author{M. Marton}
\affiliation{Laboratoire de Physique Subatomique et de Cosmologie, Universit\'e Grenoble Alpes, CNRS/IN2P3, 53, avenue des Martyrs, Grenoble, France}
\author{S. Masi}
\affiliation{Dipartimento di Fisica, Sapienza Universit\`a di Roma, Piazzale Aldo Moro 5, I-00185 Roma, Italy}
\author{A. Monfardini}
\affiliation{Institut N\'eel, CNRS and Universit\'e Grenoble Alpes, France}
    \author{M.~W. Peel}
\affiliation{Imperial College London, Blackett Lab, Prince Consort Road, London SW7 2AZ, UK}
\affiliation{Instituto de Astrofísica de Canarias, C/Vía Láctea s/n, E-38205 La Laguna, Tenerife, Spain}
\affiliation{Departamento de Astrof\'{\i}sica, Universidad de La Laguna, 
    E-38206 La Laguna, Tenerife, Spain}
	\author{G. Pisano}
\affiliation{Astronomy Instrumentation Group, University of Cardiff, UK}
\author{N. Ponthieu }
\affiliation{Univ. Grenoble Alpes, CNRS, IPAG, F-38000 Grenoble, France }
\author{R. Rebolo}
\affiliation{Instituto de Astrofísica de Canarias, C/Vía Láctea s/n, E-38205 La Laguna, Tenerife, Spain}
\affiliation{Departamento de Astrof\'{\i}sica, Universidad de La Laguna, 
    E-38206 La Laguna, Tenerife, Spain}    
    \affiliation{Consejo Superior de Investigaciones Cient\'{\i}ficas, E-28006 Madrid, Spain}
    \author{S. Roni}
\affiliation{Laboratoire de Physique Subatomique et de Cosmologie, Universit\'e Grenoble Alpes, CNRS/IN2P3, 53, avenue des Martyrs, Grenoble, France}
\author{S. Roudier}
\affiliation{Laboratoire de Physique Subatomique et de Cosmologie, Universit\'e Grenoble Alpes, CNRS/IN2P3, 53, avenue des Martyrs, Grenoble, France}
\author{J.~A. Rubi\~no-Mart\'in}
\affiliation{Instituto de Astrofísica de Canarias, C/Vía Láctea s/n, E-38205 La Laguna, Tenerife, Spain}
\affiliation{Departamento de Astrof\'{\i}sica, Universidad de La Laguna, 
    E-38206 La Laguna, Tenerife, Spain}
	\author{D. Tourres }
\affiliation{Laboratoire de Physique Subatomique et de Cosmologie, Universit\'e Grenoble Alpes, CNRS/IN2P3, 53, avenue des Martyrs, Grenoble, France}
\author{C. Tucker}
\affiliation{Astronomy Instrumentation Group, University of Cardiff, UK}
\author{T. Viera-Curvelo}
\affiliation{Instituto de Astrofísica de Canarias, C/Vía Láctea s/n, E-38205 La Laguna, Tenerife, Spain}
\affiliation{Departamento de Astrof\'{\i}sica, Universidad de La Laguna, 
    E-38206 La Laguna, Tenerife, Spain}
    \author{C. Vescovi}
\affiliation{Laboratoire de Physique Subatomique et de Cosmologie, Universit\'e Grenoble Alpes, CNRS/IN2P3, 53, avenue des Martyrs, Grenoble, France}



\begin{abstract}
   Kinetic inductance detectors (KIDs) have been proven as reliable systems for astrophysical observations, especially in the millimetre range. Their compact size enables to optimally fill the focal plane, thus boosting sensitivity.
   The KISS (KIDs Interferometric Spectral Surveyor) instrument is a millimetre camera that consists of two KID arrays of 316 pixels each coupled to a Martin-Puplett interferometer (MPI). The addition of the MPI grants the KIDs camera the ability to provide spectral information in the 100 and 300\,GHz range.
   In this paper we report the main properties of the KISS instrument and its observations. We also describe the calibration and data analysis procedures used. We present a complete model of the observed data including the sky signal and several identified systematics.
   We have developed a full photometric and spectroscopic data analysis pipeline that translates our observations into science-ready products. We show examples of the results of this pipeline on selected sources: Moon, Jupiter and Venus.
   We note the presence of a deficit of response with respect to expectations and laboratory measurements. The detectors noise level is consistent with values obtained during laboratory measurements, pointing to a sub-optimal coupling between the instrument and the telescope as the most probable origin for the problem. This deficit is large enough as to prevent the detection of galaxy clusters, which were KISS main scientific objective. Nevertheless, we have demonstrated the feasibility of this kind of instrument, in the prospect for other KID interferometers (such as the CONCERTO instrument). 
   As this regard, we have developed key instrumental technologies such as optical conception, readout electronics and raw calibration procedures, as well as, adapted data analysis procedures. 
\end{abstract}

\keywords{ instrumentation: interferometers ---   techniques: imaging spectroscopy ---  methods: observational}


\section{Introduction} \label{sec:intro}
\label{sec:introduction}

The Sunyaev-Zel'dovich effect \citep[SZe,][]{Sunyaev_Zeldovich:SZE_forecast_original_paper, Sunyaev_Zeldovich:SZE_1972}
is one of the most important and thoroughly studied secondary CMB 
anisotropies. CMB photons undergo inverse Compton scattering with
electrons in hot gas regions, which produces an increase of the photon 
energy. The electron temperature is particularly high in the intra-cluster 
medium within galaxy clusters (GCs), making the SZe a useful probe to detect
distant matter structures. Because of the independence of the SZe signal
with respect to the redshift at which each GC is located, it has been
extensively used for cosmological analysis in the past 
\citep[e.g.,][]{planck_cluster_cosmology_2013, 
planck_cluster_cosmology_2015}. 
Besides this signal, which is normally referred to as the
thermal SZe (or tSZe), an additional component of the SZe must be taken 
into account when the velocity of the GC is not negligible with respect
to the CMB reference frame. This component receives the name of kinematic
SZe (or kSZe), and it has been studied in less depth \citep{ACT_kSZ, 
planck_kSZ_w_SPT_ACT} because of its lower amplitude (and therefore, 
more challenging detection). 

New physics is available when an accurate description of the SZe beyond
its thermal component is achieved, with a special focus on kSZe 
\citep{kSZ_new_1, kSZ_new_2, kSZ_WP_1, kSZ_WP_2}. 
One can rely on their different spectral dependencies in 
order to separate the two, but precise measurements are required due to the 
smaller amplitude of kSZe. Therefore, gathering as much data as possible in 
the millimetre (100--300\,GHz) range, where SZe is most important, is
mandatory to perform an accurate separation between the two components.
Constraining the emission around 217\,GHz is especially enlightening, as 
the tSZ contribution is null at that frequency, so every shift from the
expected CMB blackbody distribution will be adscribed to the kSZe.

One way to obtain these data would be to use spectrometres in the mm-band.
Martin-Pupplet interferometers \citep[MPI,][]{mpi} allow to obtain 
interferometric data produced by a differential signal under a 
varying optical path difference (OPD). 
The obtained interferograms are later transformed into the actual
spectral energy distribution (SED) from the observed source after running
a Fourier transformation. Several other instruments have used Fourier 
Transform Spectrometers (FTS) in the past, both ground-based (e.g. EMIR 
\citealt{EMIR_IRAM} or SEPIA \citealt{SEPIA_APEX}) and onboard space 
satellites (e.g. FIRAS \citealt{FIRAS} or SPIRE \citealt{herschel_spire}). 
However, for the first time, we propose to couple such a spectrometer
with kinetic inductance detectors
\citep[KIDS,][]{kids}, while most of the previous ones used bolometric or 
heterodyne receivers. KIDs have been used already in photometric
mm-instruments such as the NIKA and NIKA2 cameras 
\citep{Monfardini_2010:NIKA, nika2}, both installed at the IRAM 30-m telescope at 
Pico Veleta. The two have provided extensive information on GCs in the mm 
band for the last 15 years \citep[e.g.,][]{firstGCobs_w_KIDS, perotto2020}.

In order to test this concept and as a testbed for the CONCERTO instrument  \citep{concerto},
we constructed and operated the KISS instrument \citep{fasano-ltd}. KISS 
was installed in the first QUIJOTE \citep{quijote2010}
telescope at the Teide Observatory in Tenerife, from 2018 to 2020. The 
instrument consisted of a KIDs camera coupled to the actual warm MPI, which is coupled to the telescope through a series of tailored optics.
The data  
discussed in this work covers a frequency range between 120 and 180\,GHz,
while data with an additional band between 200 and 300\,GHz have been 
discussed in previous works \citep[e.g.,][]{fasano-moonspectra}. The 
CONCERTO \citep{concerto} instrument followed a similar instrumental 
configuration while pursuing much finer sensitivity and angular resolution. 
It was installed in the larger APEX \citep{APEX} telescope in Chile,
with a new and customized set of KIDs detectors.

This paper is organized as follows: in Section~\ref{sec:kissexp} we introduce 
the KISS instrument and in Section~\ref{sec:dataprocessing} we discuss how 
the raw data is transformed into scientific products. We explain the
characteristics and systematics of photometric and spectroscopic data in 
Sections~\ref{sec:photometry} and \ref{sec:spectroscopy}, respectively. 
We finally present the scientific results on astrophysical targets in 
Section~\ref{sec:specastro}, and the conclusion remarks in 
Section~\ref{sec:conclu}.
\section{The KISS instrument}
\label{sec:kissexp}
KISS is a ground-based millimetre Fourier transform spectral imager installed at the first QUIJOTE telescope \citep{QUIJOTE, 10.1117/12.926416}, one of the two 2.25\,m telescopes of the QUIJOTE experiment at the Teide Observatory\footnote{Located at 2\,400\,m of altitude in Tenerife, Spain.}. Operations ran from 2018 to 2020. KISS is based on a Martin-Puplett interferometer \citep[MPI,][]{mpi} coupled to the telescope and to a millimetre camera made of two arrays of KIDs operating in the frequency range from 120 to 180\,GHz. KISS has been designed to achieve a typical angular resolution of 7 arcmin at 150\,GHz with a field-of-view (FoV) of 1 degree. The KISS performance parameters are summarized in Table~\ref{table:basic_parameters}. 

\begin{table}
\centering
\caption{Basic performance parameters of KISS. We find a lower value when using the Moon as a calibrator, as explained in Section~\ref{sec:moon_photometry}. The calibration factor obtained from spectral data is also lower than that obtained from photometry, as discussed in Section~\ref{sec:spectroscopy}. \label{table:basic_parameters}}   
 \begin{tabular}{cc}
       \hline \hline
       Parameter & Value \\ \hline
       Total number of KIDs & 632  \\
       Central frequency (GHz)  & 150 \\
       Bandwidth (GHz)  & 60 \\ 
       Primary mirror diametre (m) & 2.25 \\
       Field-of-view (FoV, degrees) & 1 \\
       Sampling frequency (kHz)  & 3.816  \\
       \hline
       Beam FWHM 150\,GHz (arcminutes) & 7 \\
       Beam ellipticity, $e$ (arrays A \& B) & 0.175--0.275 \\
       \multirow{2}{*}{White noise level @ 150\,GHz} & 1.5--2\,Hz\,s$^{1/2}$ \\
         & 60--80\,mK\,s$^{1/2}$ \\
       $NEFD$ @ 150\,GHz (\,Jy\,s$^{1/2}$) & 140--180\\
       Calibration factor from & \multirow{2}{*}{$35.7\pm2.6$}\\
       photometry on Jupiter (Hz\,K$^{-1}$) & \\
       \hline
       Maximum mirror displacement (mm) & 30 \\
       Interferogram rates (Hz) & 7.45 \\
       Spectral Frequency range (GHz) & 120--180 \\
       Results spectral binning (GHz) & 8.33 \\
       \hline
    \end{tabular}
     
\end{table}
%

\begin{figure*}[h]
\begin{center}
\gridline{\fig{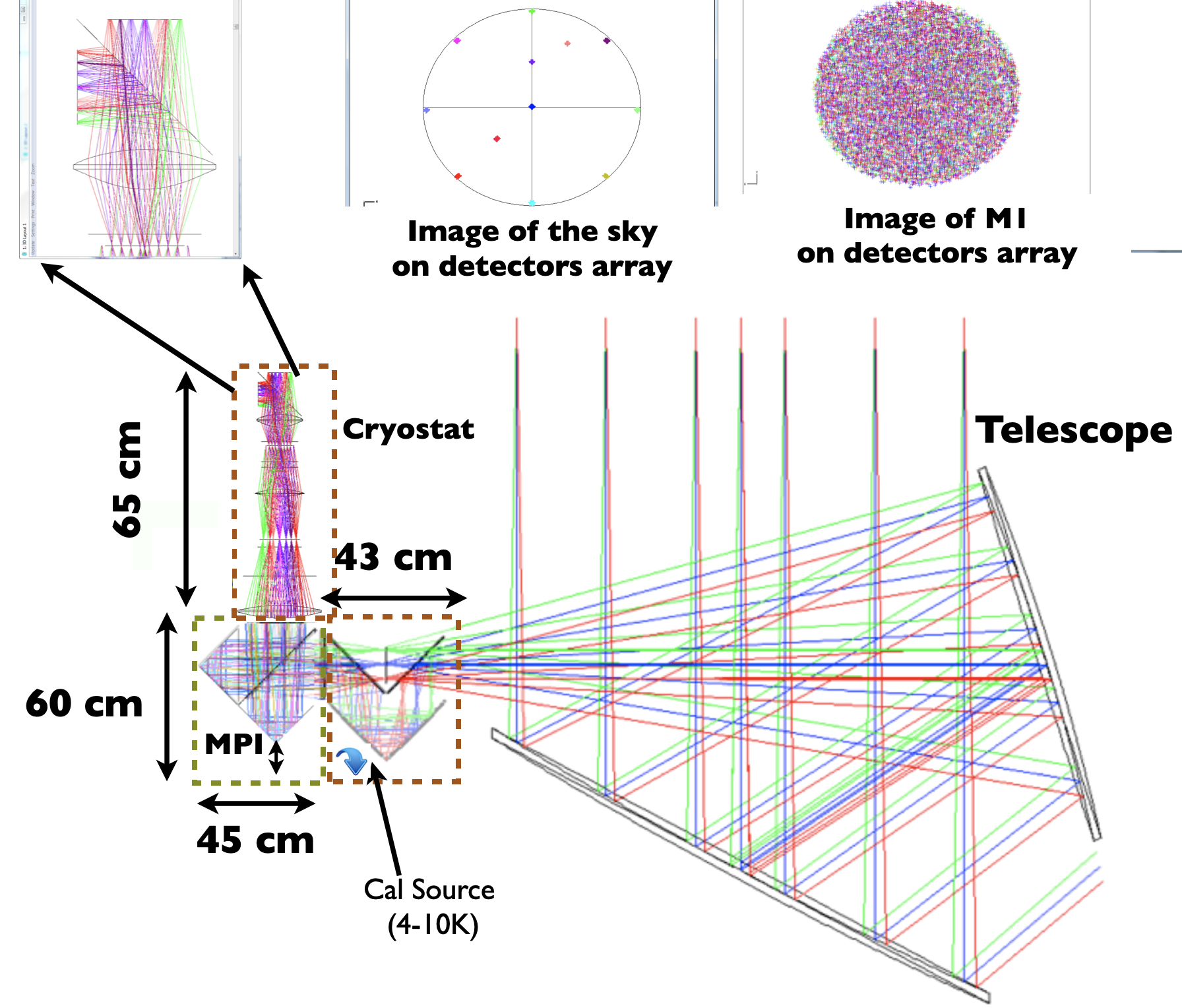}{0.5\textwidth}{(a)}, \fig{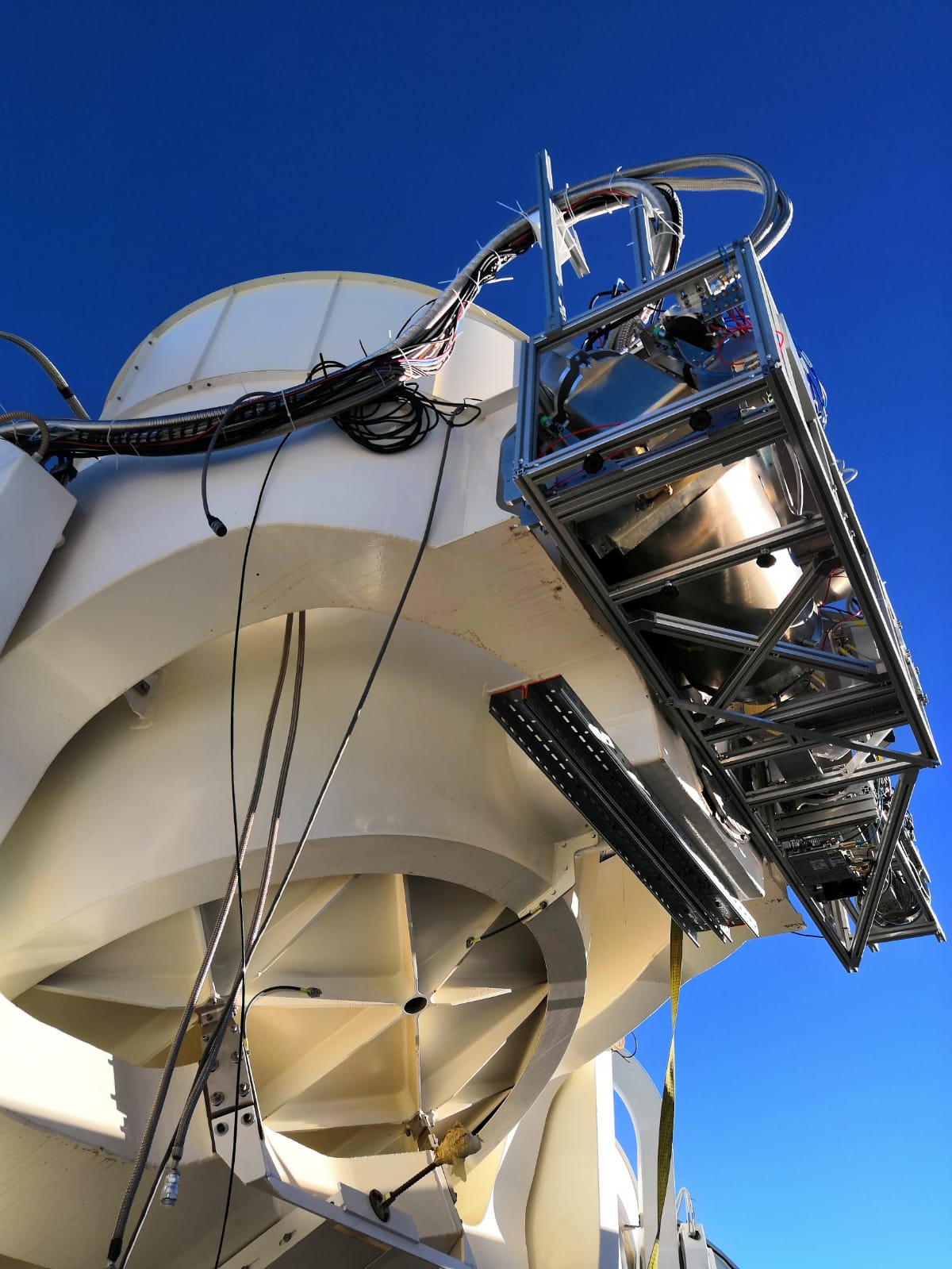}{0.3\textwidth}{(b)}}
\gridline{\fig{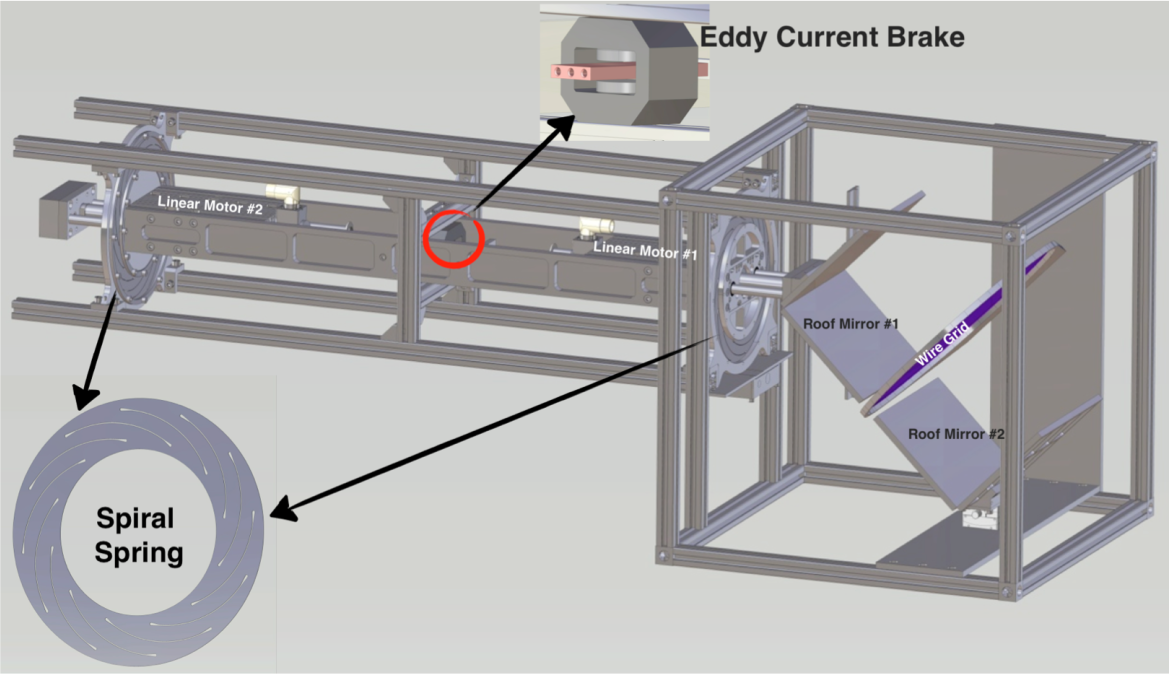}{0.4\textwidth}{(c)},\fig{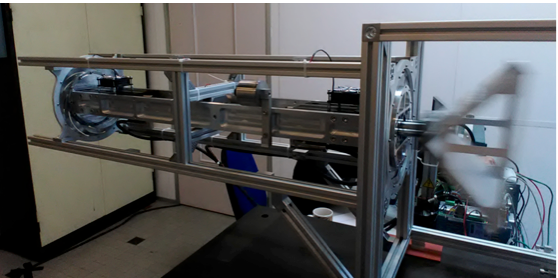}{0.4\textwidth}{(d)}}
\caption{The fast MPI interferometer developed for the KISS experiment. (a): the optical scheme of the KISS instrument. (b): Picture of the instrument installed at the QUIJOTE telescope. (c) and (d): comparison between a mechanical drawing and the real KISS Martin-Puplett Interferometer. It consists of two roof mirrors, from which one of them can move up to 100\,mm with a frequency of 5\,Hz. \label{MPI}}
\end{center}
\end{figure*}
\begin{figure}[h]
	\centering
	\fig{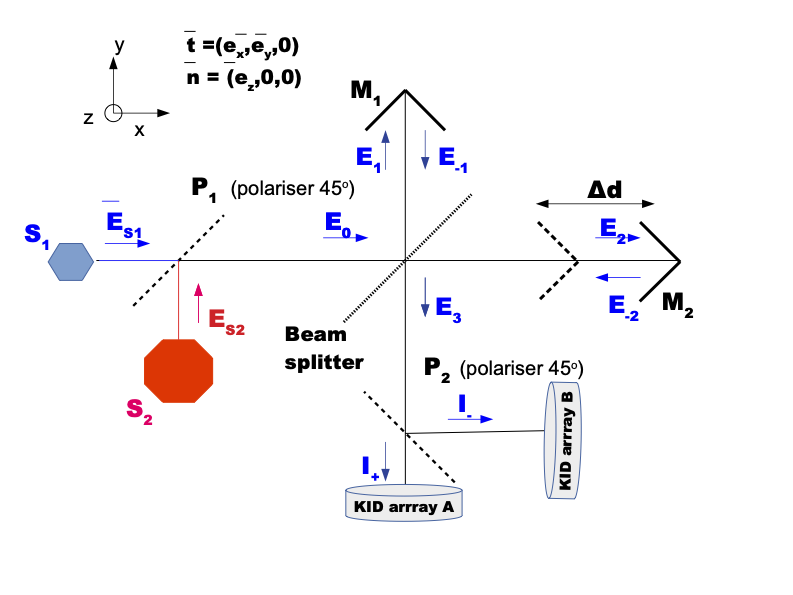}{0.5\textwidth}{}
	\caption{KISS Martin-Puplett schematic view. Input signals $S_1$ and $S_2$ are combined to produce an interference pattern thanks to the displacement of a roof mirror (M2) with respect to a fixed one (M1). \label{fig:kissmpiconcept} }
\end{figure}
\begin{figure}[h]
	\centering
 	\fig{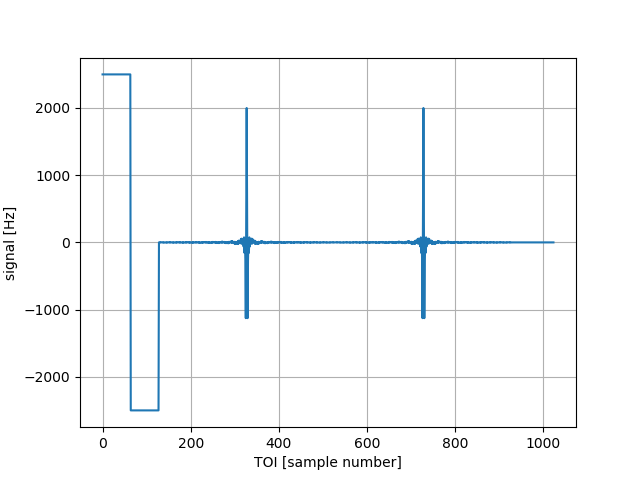}{.5\textwidth}{}
	\caption{Simulated KISS data block as presented in \cite{fasano-calib}. The first, squared-shaped samples correspond to the modulation used for the 3-point calibration, which is described in depth in Section 2 of \cite{fasano-calib}. Then, we observe the \fwd\ (left/first) and \bwd\ (right/second) interferograms. \label{fig:intertotal}}
\end{figure}


\subsection{The KISS Martin-Puplett Interferometer}
\label{sec:KISS_MPI}
A MPI measures the difference between two optical inputs, as all FTS, but in this case the beam is split in two with different linear polarization (\citealt{mpi}, see Fig.~\ref{fig:intertotal}). This can be done with the use of two initial wire-grid polarisers. After interfering with the MPI, the first half of the beam is focused onto the detectors, creating an image of 1 deg on the sky. The second half of the beam is first defocused from a focused image to a pupil (corresponding to the image of the primary mirror of the telescope) and then collected by the detectors. This accounts for a constant signal over the whole FoV. From the point of view of the spectroscopic measurements, this is equivalent to an initial hardware subtraction of the constant common mode from the atmosphere (which simplifies the offline analysis)
Instead of the sky itself, this second, flat-input can be taken also from a cold reference, which can be either the detectors themselves by $narcissus$ effect (i.e. being reflected by a mirror - \citealp{Howard:82}) or a cold black-body at a given tuneable temperature.
The KISS MPI posed a technological challenge due to the fast scanning speed required to overcome the effects of large scale atmospheric fluctuations. 
The control of the systematic errors of such a configuration drives the major requirement for the whole instrument. The MPI must be able to perform continuously 3\,cm path interferograms (which give a spectral resolution of about 3\,GHz within the band) at a rate of 7.45 Hz\footnote{Corresponding to two interferograms per block of 1024 samples with a sampling rate of 3.816\,kHz.} to maintain a low sky noise level and sufficient angular sampling on the sky. The fast KID response time permits us to use this method without any loss of information. This requirement also constrains the readout sampling frequency. We found that a few hundred data points for each interferogram are needed in order to preserve all the spectral information from the data in the band of interest. This means that the readout electronics have to operate at a sampling frequency of 3.816\,kHz. \\
These requirements have been studied and a solution is presented in Fig.~\ref{MPI}. One of the two roof mirrors is moved by a mechanical-bearing direct-drive linear stage allowing for backward and forward displacement at 3.725\,Hz. 
In order to achieve accurate linear movement and reduce vibration effects, which would significantly increase the noise in the data, we used a flat spiral spring similar to that developed for the Planck HFI 4\,K compressor. In this geometry, the flexing movement is produced by arms that connect inner and outer rings \citep{fasano-ltd}. Flexing allows the planes defined by these rings to move easily with respect to each other by rotation perpendicular to the flexing axis, displacement, or a combination of both. 
\subsection{The KISS camera and readout electronics}
The KISS camera design is based on the NIKA and NIKA2 cameras \citep{Monfardini_2010:NIKA,nika2} sharing the same KID technology \citep{Swenson} and readout electronics \citep{2012JInst...7.7014B,2012SPIE.8452E..0OB,2013JInst...8C2006B,2016JInst..1111101B}. The optical coupling between the camera, the MPI and the telescope is ensured by a series of lenses (see Fig.~\ref{MPI}.
The 1 deg focal plane is fully covered by two arrays of single polarisation aluminium lumped-element KID (LEKID) containing 316 pixels each and illuminated by a 45 degrees folded polariser. Each array has a common optical band 120--180\,GHz. KISS detectors are operated at a temperature of 170\,mK, stabilized at 0.5\,mK for elevation change. This operation temperature is obtained via a $^{3}$He-$^{4}$He dilution cryostat based on the NIKA camera cryostat \citep{Monfardini_2010:NIKA}, which has been adapted for KISS. \\ \\
The KISS detectors are instrumented by two (one per array) AMC NIKEL electronic boards  \citep{2016JInst..1111101B}. 
For each array, the KIDs are coupled to a single transmission line and are read simultaneously in the Fourier domain. Variations in the sky signal produce shifts in the KID resonance frequency that are measured by the readout electronics. For each KID, noted $k$, a frequency tone is injected. The electronic readout monitor then allows to recover the in-phase $(I_{k})$ and quadrature $(Q_{k})$ response of the KID to that input tone. These two quantities can be related to the KID resonance frequency shift. For this we modulate the frequency tone for each KID and apply the 3-point calibration algorithm described in detail in~\cite{fasano-calib}. As explained later in Section~\ref{sec:dataprocessing}, KISS was operated in sky-sky mode for which we expect moderate signal amplitudes and as so in accordance with the 3-point calibration procedure. In this was not the case an extra modulation cycle at the end of each scan can be introduced to reconstruct the resonance shape for each detector. This has been already considered in the updated version of KISS software used in CONCERTO \citep{2024arXiv240615572H}.

\section{Raw data and observation modes}
\label{sec:dataprocessing}
The KISS instrument can be considered as a spectral imager, allowing us to obtain spectral and photometric data simultaneously over the full FoV. For a given scanning strategy we intend to reconstruct the sky spectral energy distribution (SED) for each sky position. As for other MPI based-instruments SEDs can be reconstructed directly from the Fourier transform of interferograms, as detailed in the following section. These interferograms are produced by displacing the movable roof mirror from a given initial position to the maximum displacement and back to the initial position, as introduced in Section~\ref{sec:kissexp}. 
Fig.~\ref{fig:kissmpiconcept} shows a schematic view of the KISS MPI and the interface with the telescope and the KIDs arrays.
\subsection{Interferograms and spectral energy density}
Considering two multi-chromatic input sky signals $S_{1}(\nu)$ and $S_{2}(\nu)$ we can write the total observed intensity in the two detector arrays as:
\begin{equation}
    I_{\pm}  = S^{phot} +  I^{sp}_{\pm}
\label{eq:total_observed_intensity}
\end{equation}
with
\begin{small}
\begin{eqnarray}
&S^{phot} = \int_{-\infty}^{+\infty} d\nu \ H (\nu) \ T^{atm}(\nu) \left( \frac{S_{1}(\nu)+ S_{2}(\nu)}{4} \right),  \\
&I^{sp}_{\pm} = \pm  \int_{-\infty}^{+\infty} d\nu\ H (\nu)  T^{atm}(\nu) \left( \frac{S_{1}(\nu)-S_{2}(\nu)}{4} \right)  \cos{ \left( 2 \pi \nu \frac{2 \Delta d(t)}{c} \right) },
\label{eq:kissinter}
\end{eqnarray}
\end{small}
\noindent where $H (\nu)$ and $T^{atm}(\nu)$ represent the KISS frequency transmission and the atmospheric transmission, respectively. The relations between the total observed intensity from Eq.~\ref{eq:total_observed_intensity} and the electrical fields propagating through the instrument (as shown in Fig.~\ref{fig:kissmpiconcept}) are detailed in Appendix~\ref{appendix:interferometric_derivation}. The $\pm$ relates to the in-transmission or in-reflection outputs of the MPI corresponding to KID arrays A and B, as shown in Fig.~\ref{fig:kissmpiconcept}. \\ \\\
The first term can be considered as the total photometric continuum contribution including both input signals: $S_{1}$ and $S_{2}$. The second term can be identified as the spectroscopic measurement. In practice, as discussed above, by moving the roof mirror in the range $-\Delta d_{max}$ to $\Delta d_{max}$ it is possible to reconstruct the SED of the input signal as the inverse Fourier transform of the interferometric signal:
\begin{equation} 
S_{1} (\nu) - S_{2} (\nu) =  \frac{\mathcal{R} \left[ \int_{-\Delta d_{max}}^{+\Delta d_{max}} dx \ e^{2 \pi \nu \frac{x}{c}} I^{sp}_{\pm}(x) \right] }{H (\nu) T^{atm} (\nu)}
\label{eq:kissspectro}
\end{equation}
where $\mathcal{R}$ represents the real part.
\begin{figure*}[h]
	\centering
	\fig{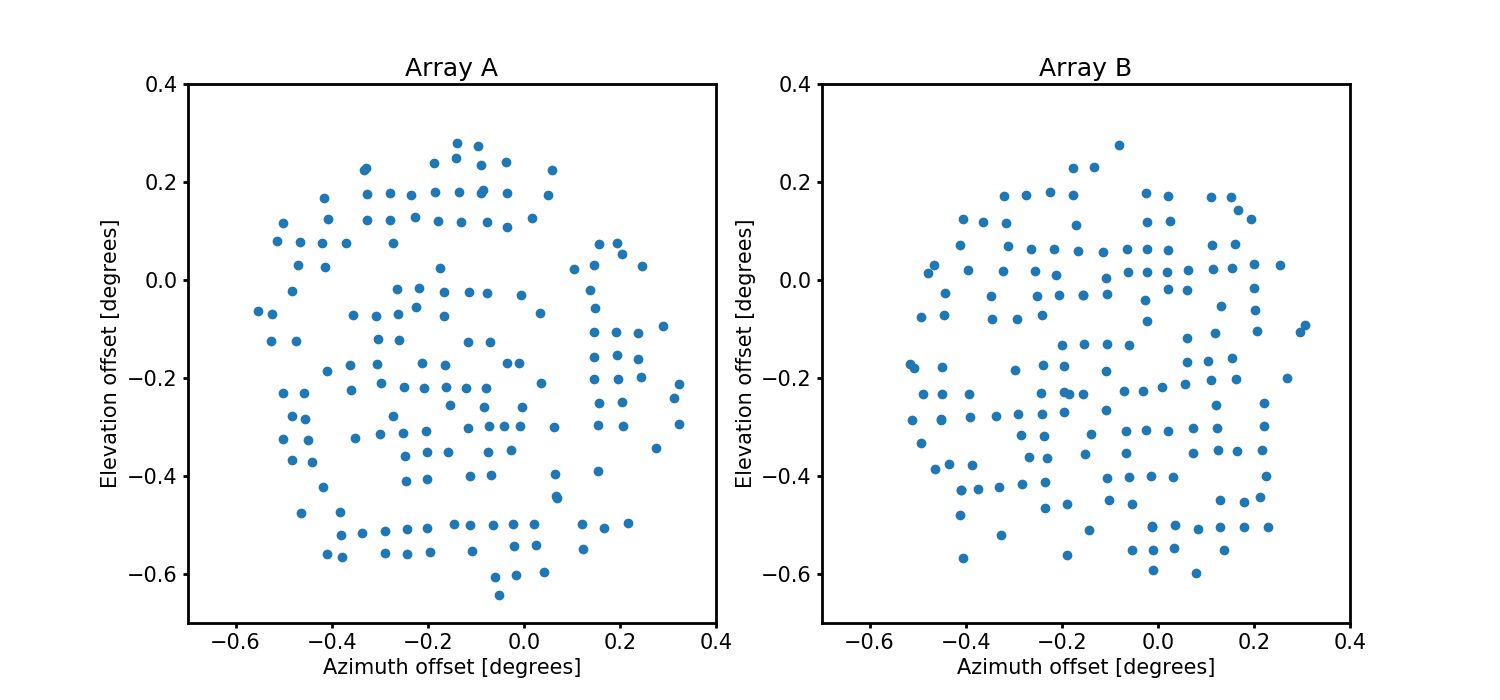}{\textwidth}{}
	\caption{On-sky focal plane geometry. KID offset position for array A (left) and B (right) as computed from the median of available beammap scans on the Moon. The extended nature of the Moon, together with the joint analysis of scans taken at different elevations, explain the variable spacing between detectors (opposed to the regular spacing obtained from laboratory measurements in \ref{fig:labfocalplane}). We observe some empty regions in the focal plane corresponding to KIDs that were not selected for the final analysis.} 
	\label{fig:skyfocalplane}
\end{figure*}
\begin{figure}[h]
	\centering
	\fig{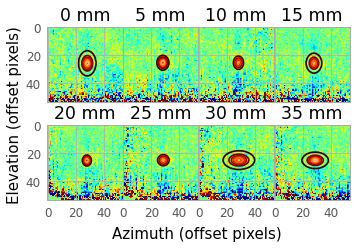}{0.45\textwidth}{}
	\fig{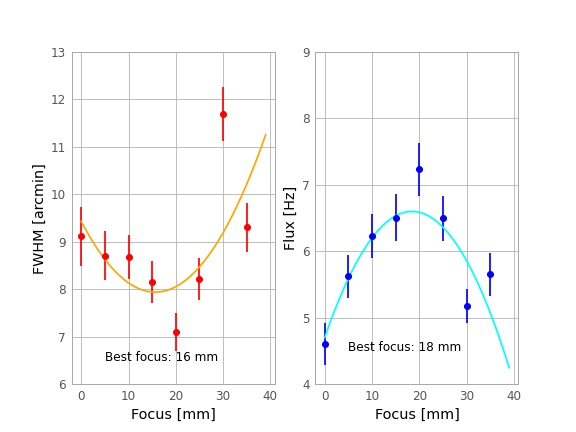}{0.5\textwidth}{}
	\caption{Focus. Top: Jupiter maps at 150\,GHz obtained from the combination of all detectors for different focus distances. Low (high) values return prolate (oblate) morphologies. Bottom: changes in the maximum FWHM of the disks and the integrated flux with variable focus distance.}
	\label{fig:phfocus}
\end{figure}
\subsection{Raw signal data and modulation}
\label{section:description_raw_signal}
The raw signal data from the KISS instrument consist of blocks of 1024 points at a sampling rate of 3.816\,kHz\footnote{This means that, in reality, less than 4 blocks (3.816) are registered each second.}. For each sample we register $(I_{k},Q_{k})$ for each KID. The readout electronics and MPI motors are synchronized so that the roof mirror does a full forward-backward displacement producing a \fwd\ ($-\Delta d_{max}$) and a \bwd\ ($+\Delta d_{max}$) interferograms within each block. The initial position of the displacement mirror is set so that the zero path difference (ZPD) is well-centered with respect to the mirror path. Two lasers allow us to monitor the displacement of the mirror and thus to reconstruct the 
optical path difference (OPD). \\

In order to define the optimal MPI sampling frequency, we first assumed a $1/f$ spectrum for the atmospheric contribution with a high knee frequency of 1\,Hz. This was taken as an extreme case from the NIKA2 instrument \citep{perotto2020}. Then, the sampling frequency we set to its maximum value (1\,kHz), and the MPI frequency is set to the value that maximizes the number of acquired interferograms while maintaining a low atmospheric noise contribution. This is naturally obtained by pushing the interferometric signal to frequencies above the expected atmospheric knee frequency. \\

In practice, the modulation of the frequency tone is performed during the first 128 samples of each block, considering 64 samples with positive modulation and 64 samples with negative modulation. These modulations are used to calibrate the response from the KID. The other 896 samples are used to obtain the two interferograms, which leads to 448 samples for each of them. As the interferogram is centered with respect to the ZPD, the maximum number of sampled points for the signal is 224. A typical block of data is shown in Fig.~\ref{fig:intertotal}. 
In the following, we define  $N_{spb} = 1024$ as the number of samples per full spectroscopic cycle or block; $N_{mod}  = 128$ as the number of modulated samples, and $N_{int} = 448$ as the number of data samples per interferogram. Using the 3-point calibration algorithm described in \cite{fasano-calib} we can compute photometric and spectroscopic estimates of the signal as described in the following sections.
\subsection{Observation Mode}
One of the objectives of KISS was to prove that atmospheric contamination can be reduced in real time.
For this, as discussed above, during astrophysical observations we decided to set the second input as the constant signal along the focal plane. Assuming that the sky signal is given by the combination of a compact (with respect to the FoV) astrophysical source and the atmospheric emission, we can write:
\begin{small}
\begin{eqnarray}
\left( S_{1}(\nu)+ S_{2}(\nu) \right)(\Omega) =  \int  d\Omega^\prime& \ B_{\nu} (\Omega^\prime-\Omega) \  S^{\mathrm{source}}(\nu)(\Omega^\prime) \\
& + 2*B^{atm}(\nu)+\delta B^{atm}(\nu)  \\
\left( S_{1}(\nu)-S_{2}(\nu) \right)(\Omega)   =  \int  d\Omega^\prime& \ B_{\nu}(\Omega^\prime-\Omega) \  
\label{eq:signalbackground}
\end{eqnarray}
\end{small}
where $B_{\nu}$ is the beam pattern and $\Omega$ is the observed position on the sky.
$S^{\mathrm{source}}(\nu)$ represents the spectrum of the source. $B^{atm}(\nu)$ is the averaged atmospheric background signal and $\delta B^{atm}(\nu)$ atmospheric spatial fluctuations across the FoV. Notice that in these equations we did not account for the displacement of the source during the time of observation of a block.
\section{Continuum analysis and calibration}
\label{sec:photometry}
\subsection{Continuum signal reconstruction and map--making}
\label{sec:photoreconstruction}
Following the discussion in Sect.~\ref{sec:dataprocessing} the continuum signal for the $k$-th KID as a function of time 
can be expressed to first order approximation as:
\begin{eqnarray}
S^{phot}_{k}(t) = C^{k} \ S^{phot}_{k} (t_{b})+ A^{atm}_{k} \left( 2B^{atm}(t)+ \delta B^{atm}(t) \right) & \\
+ A^{elec}_{k} \ n^{elec}(t) + & n_{k}(t) \\
\label{eq:continuum_analysis}
\end{eqnarray}
where $C_{k}$ is a calibration factor per KID accounting both for response variations between detectors and a global calibration factor. $n_{k}(t)$ is the detector noise contribution, and $n^{elec}(t)$ represents the detector correlated electronic noise. We assume the atmospheric and electronic correlated noise between detectors with scaling coefficients per KID $A^{atm}_{k}$ and  $A^{elec}_{k}$, respectively. Thus, to correct for the atmospheric and electronic noise we subtract a baseline and perform a standard decorrelation analysis. We construct a common mode template of the KID signals per array (or per electronic box) and per subscan (regions with constant azimuth or elevation within the scan). In practice, we find that the common signal is dominated by an electronic noise component induced by 50\,Hz pickup. With a linear regression analysis, we compute the contribution of this common signal to the data of each KID and subtract it, as well as a constant value per subscan. These corrections are similar to those applied also for spectral data, as further explained in Section~\ref{sec:spectroscopy}. The corrected data are then projected into sky maps. In the case of very bright sources (such as the Moon) a simple median baseline per subscan is subtracted.
\subsection{Focal plane geometry reconstruction}
\label{sec:fpgeom}
As it is later discused on Section~\ref{sec:continuum_sensitivity}, we found the response of the instrument to be severely lower than expected from laboratory measurements (a 30--40 factor worse). This is expected to be due to the anomalous optical coupling between the QUIJOTE telescope and the KISS camera. This low response implies that even the brightest planets (Jupiter and Venus) are too faint to derive a reliable focal plane geometry (i.e. none of the two are directly visible from individual detector maps)\footnote{In addition, planet measurements were used later to derive the overall beam properties, after co-adding all detectors for the two arrays. These properties are described in the following Section~\ref{sec:focopt}, and are the ones quoted in the summary Table~\ref{table:basic_parameters}.}. Thus, we used Moon observations to derive the position of each KID in the focal plane and their relative response. We performed a large number of raster scans of the Moon and selected the best ones in terms of weather conditions and instrumental behavior. For each scan, we produced photometric maps per detector and fit the Moon signal on those to both a 2D Gaussian beam and to a model of the Moon signal (mainly an uniform disk of 30$^{\prime}$ diameter) convolved by the expected beam. Overall we obtained similar results but those related to the 2D Gaussian fit proved to be more robust in terms of KID position. The recovered KIDs positions in the focal plane for the whole set of scans are shown in Fig.~\ref{fig:skyfocalplane} for array A and B. The average from the best scans was used to obtain the final geometry of the focal plane. The results are consistent with the laboratory measurements presented in Appendix~\ref{appendix:lab_measurements}.
\subsection{Focus optimization}
\label{sec:focopt}
An accurate characterization of the focal distance of the instrument is required in order to prevent a significant degradation of the properties of its beam. This would be translated into an increase of its ellipticity and FWHM, both would also imply a loss of the integrated flux (assuming in both cases a resolved source). The position of the focal plane with respect to the optical system can be changed in order to calibrate the optimal focus position (or focal distance). This corresponds to 20\,mm with respect to the dedicated camera mount on the telescope. In the top plots of Fig.~\ref{fig:phfocus} we show maps of Jupiter performed at different focus positions. By fitting the Jupiter signal to a 2D circular Gaussian we obtain the FHWM and amplitude as shown in the bottom panels. The lowest FWHM values and flux losses are achieved for focal distances, with respect the telescope mount, between 16 and 18\,mm, respectively. The results presented in this paper were obtained for a distance of 18\,mm with respect to the telescope mount. We measured little to no variations in the beam properties with elevation, and neither did the KISS sensitivity degradation due to the sub-optimal QUIJOTE/KISS coupling (as presented below) degrade the beam properties. Therefore, we used an average focus offset along all observable elevation angles (30--90\,deg).
\subsubsection{Absolute continuum calibration}
\label{sec:absphcal}
\begin{figure}[h]
	\centering
	\fig{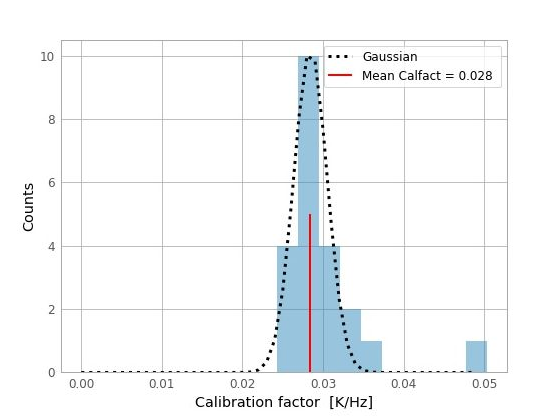}{.5\textwidth}{}
	\caption{Distribution of the joint continuum calibration factors for the A and B KID arrays obtained from the selected Jupiter measurements. The mean value and uncertainties presented in the figure are considered for the results presented in this paper. }
	\label{fig:phabscal}
\end{figure}

We used Jupiter as the main KISS continuum calibrator. 
The Jupiter brightness temperature used here is derived from the ESA 1 model \citep{planckplanets} and is equal to $170 \pm 8$\,K at 150\,GHz. All Jupiter scans showing a reduced level of noise residuals have been considered, returning consistent values independent from the atmospheric opacity or the elevation, and between the A and B KID arrays. In Fig.~\ref{fig:phabscal} we show the distribution of the calibration factors obtained for all the scans considered. This distribution is consistent with a Gaussian distribution with mean value of $35.7\pm2.6$\,Hz\,K$^{-1}$. Absolute uncertainties on the model of the brightness temperature are expected to be of the order of 5\% and are accounted for in the following.
\subsection{Continuum sensitivity}
\label{sec:continuum_sensitivity}
To obtain the final noise level in the KISS observations we have directly computed the white noise level in the TOIs and neglected the contribution from correlated atmospheric and electronic noise, which could lead to residual noise in the maps.
We obtain a white noise level ranging between $1.5$--$2.5$\,Hz\,s$^{1/2}$. This value is compatible with the detector noise measured in the laboratory and indicates that the detector behaves as expected. 
When accounting for the calibration value from Section~\ref{sec:absphcal}, this white noise level translates to $60$--$80$\,mK\,s$^{1/2}$ and a $NEFD=140$--$180$\,Jy\,s$^{1/2}$, after multiplying by the KISS solid angle. This is a large value compared to those from NIKA2 (9\,mJy\,s$^{1/2}$, \citealt{perotto2020}) or CONCERTO ($\sim$70\,mJy\,s$^{1/2}$, \citealp{2024arXiv240615572H}). This low overall response from on-telescope data
is expected to be due to the sub-optimal coupling between the instrument and the telescope. Indeed, it was not observed during the laboratory characterization
of the former, when calibration factors of $\sim1000$\,Hz\,K$^{-1}$ were 
obtained. In Section~\ref{sec:moon_photometry} we discuss how the ratio between these laboratory factors and the ones obtained from on-telescope data is around 30. When correcting for this degradation factor together with the larger beam of KISS, the results are consistent with the previously mentioned NIKA and CONCERTO estimates.
\begin{figure*}
    \centering
    \includegraphics[trim={0 20.5cm 0 0}, clip, width=1\linewidth]{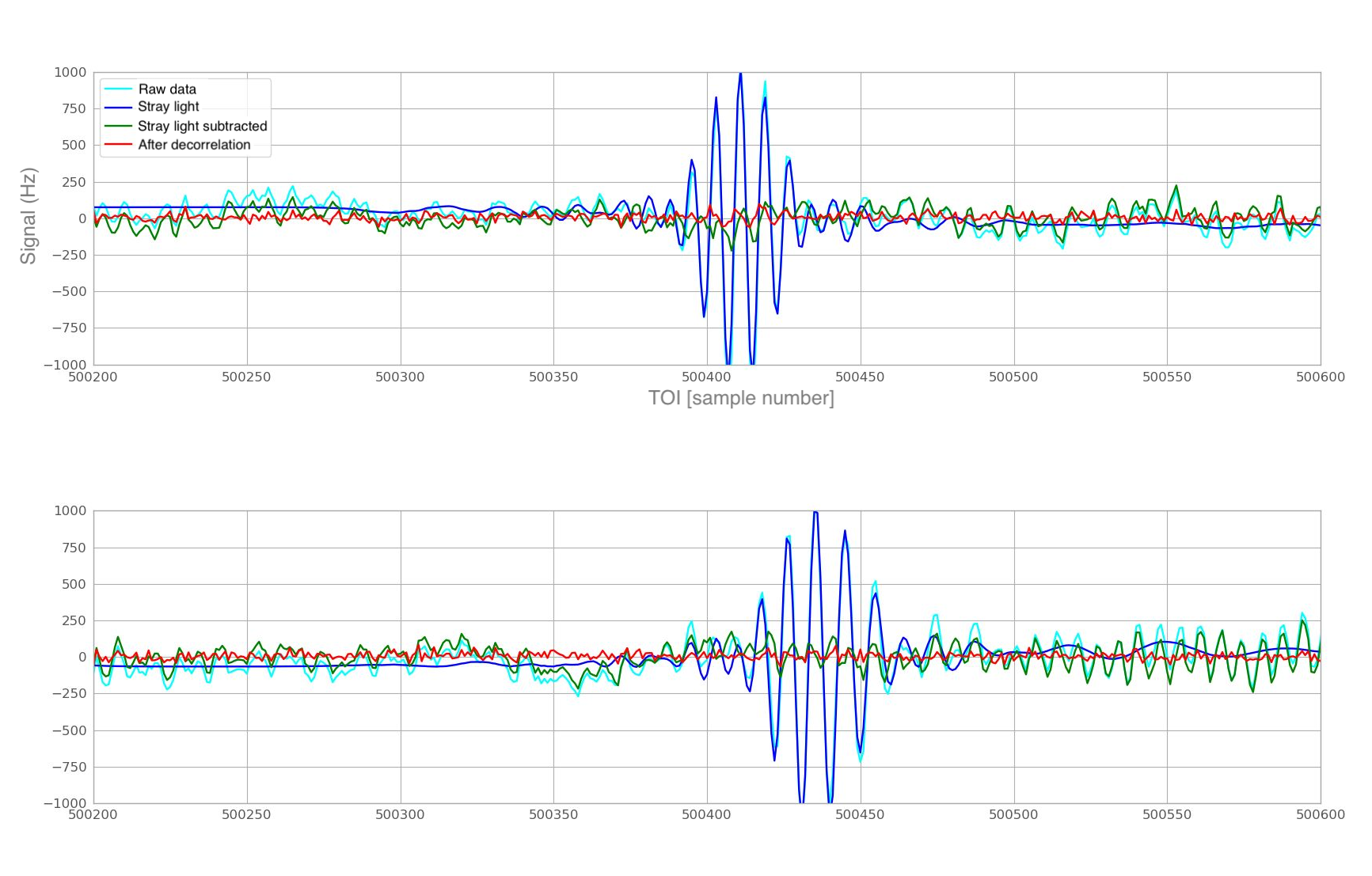}
    \includegraphics[width=0.99\linewidth]{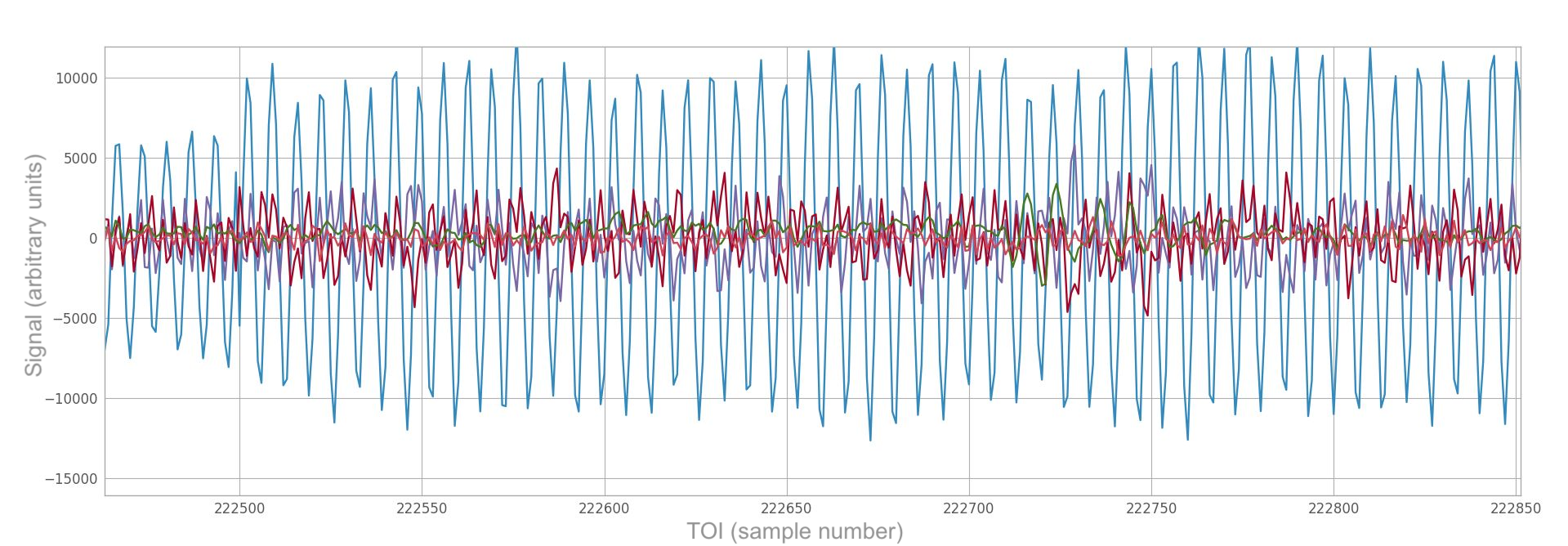}
    \caption{Top: example of KISS data after each of the steps of the data processing. The cyan line represents the raw, acquired data, while the dark blue line represents the mean stray light contribution  that effectively dominates the signal. The difference between the two is shown as the green line. The dominant component at this stage of the data processing is the 50\,Hz electrical pickup signal. In order to get rid of this signal (and other residual ones) a PCA is performed to estimate the noise behaviour. The final decorrelated signal, which will be projected to the maps and used for scientific purposes, is shown in red. Bottom: PCA components associated to the five most significant eigen values. These are used as template to correct for the systematic contribution discussed above (the data are assumed to be a linear combination of those templates). This correction is done independently for each KISS scan.}
    \label{fig:kiss_decorr}
\end{figure*}
 \begin{figure}
 \centering
 \fig{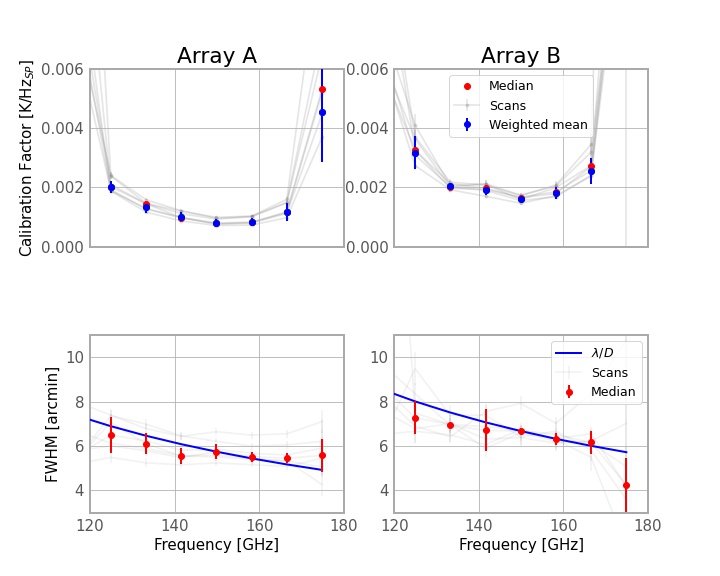}{1\linewidth}{}
\caption{Spectroscopic calibration coefficient (top) and beam FWHM (bottom) per frequency as derived from Jupiter observations for the A (left) and B (right) arrays. The calibration factors are computed as in Section~\ref{sec:absphcal}, through aperture photometry at each of the frequencies considered.}
\label{fig:spabscal}
\end{figure}
\section{Spectroscopic analysis and calibration}
\label{sec:spectroscopy}
\subsection{Data model}
\subsubsection*{Sky signal}
Starting from the signal model discussed in Section~\ref{sec:dataprocessing} we now describe the general data model of the measured interferograms for each detector. Based on Eq.~\ref{eq:signalbackground} we can write:
\begin{eqnarray}
  I_{\pm}^{\rm sp} = \pm  &\alpha_k\int_{-\infty}^{+\infty} d\nu H (\nu)  T^{atm}(\nu)   \cos{ \left( 2 \pi \nu \frac{2 \Delta d(t)}{c} \right) }\\ 
  &(A^{k}_{t,p} S_{p}(\nu) - \langle A^{k}_{t,p} S_{p}(\nu) \rangle_{FoV}) 
  + \delta B^{atm}_{p}(t,\nu)],
\label{eq:I_plusminus_obvious}
\end{eqnarray}
where the $\pm$ subscript reflects the sign difference between KIDs in the A (in-transmission) and B (in-reflection) arrays.
\subsubsection*{Detector and electronic noise}
Laboratory dark measurements have shown that the KIDs intrinsic noise can be considered white to first order. Furthermore, the analyses of NIKA and NIKA2 data \citep{perotto2020} have shown that the detected Gaussian detector correlated noise is related to the acquisition readout. In the following, the detector and electronic noise components will be noted as: 
\begin{equation}
n^{k}_{w}(t) + \rho_{k} n^{elec}(t).
\label{eq:detnoise}
\end{equation}
similarly to the components already present in Eq.~\ref{eq:continuum_analysis}. During operations, we identified a rather constant sinusoidal electrical signal that induces an additional noise component in the KISS detectors. As shown in Fig.~\ref{fig:kiss_decorr}, this sinusoidal signal has a typical period of 8 to 12 samples at 3.816\,kHz and its amplitude is modulated with time. The exact origin of this signal is not known, but it seems to be related to the electrical installation of the telescope and its coupling with the instrument. In the following, we would refer to this signal as the 50\,Hz pickup, $\rho^{50 \mathrm{Hz}}_{k} N^{50 \mathrm{Hz}}(t)$, as it seems to be an harmonic of 50\,Hz.
\subsubsection*{Parasitic mirror signal}
We identified a parasitic signal for some of the detectors that can be related to a modulation of the background between \fwd\ and \bwd\ interferograms. This is expected to be due to stray light entering the optical system when moving the displacement mirror. Therefore, it should be periodic with respect to the mirror position and we will denote it as $B^{k}_{M}(t)$. 
\subsubsection*{Stray-light background}
We have found that the KISS interferograms present a constant background signal that seems to be from environment stray-light, $I^{k}_{sl}$. This signal fully dominates the observed interferograms and needs to be removed prior to the sky projection. The background signal is expected to be consistent with a Rayleigh-Jeans spectrum at room temperature. 
\subsubsection*{Overall model}
After coadding all the components the interferometric signal measured by a detector $k$ can be written as:
\begin{eqnarray}
I^{k}_{\pm}(t) = &C^{k} \ \left[ I_{\pm}^{source} + I_{\pm}^{\delta B} \right] + \\
  &+  n^{k}_{w}(t)+ \rho_{k} n^{ele}(t) + \rho^{50 \mathrm{Hz}}_{k} N^{50 \mathrm{Hz}}(t) +B^{k}_{M} (t) + I^{k}_{sl} \\
\label{eq:kissdatamodel}
\end{eqnarray}
where the first two terms account for Eq.~\ref{eq:I_plusminus_obvious}, the former for the difference between the inputs and the latter for the atmoshere fluctuations. 
\subsubsection*{Other possible systematic effects}
Other systematic effects can affect the reconstruction of the spectrum for a generic FTS. In the case of KISS, the lack of instrumental response at the telescope prevented us from studying those in detail. We just mention here the most important ones and comment on their mitigation during instrumental design and laboratory testing. An important concern during design process was to ensure an homogeneous illumination of the focal plane and in particular of off-axis pixels. The shape of the KISS mirrors and lenses were optimized accordingly and we slightly under-illuminated the telescope. Another important systematic effect comes from difference in the optical path or misalignment of the two arms of the interferometer. To correct for this we extensively performed alignment campaigns both during test in the laboratory and prior to telescope installation. The alignment was considered optimal when maximizing the amplitude of the interferograms for all detectors. We observed no hint of misalignment on the sky data. We have also considered variations in the optical path induced either by vibrations of the mechanical system or of the beam splitter. For the former, as discussed above, we considered a spiral spring and a two opposite motor system that made mechanical vibrations contribution to the data negligible. Vibrations of the beam splitter can produce inhomogeneous time variations of the optical path across the field of view and lead to differences on the sampling of the OPD (not in accordance to the measurement from the laser monitoring) and on the position of the ZPD. For KISS we only observed time  constant variations of the ZPD position across the field-of-view as discussed above. Finally, we have observed variations between the OPD sampling for the \fwd\ and \bwd\ interferograms but those can be corrected using the laser monitoring.

\subsection{Spectroscopic signal reconstruction and map making}
As shown in Eq.~\ref{eq:kissinter}, after the 3-point calibration has been performed we construct independent arrays containing the \fwd\ and \bwd\ interferograms for each detector and for each block. For the sake of clarity, we express the different contributions to the raw interferograms described above in terms of detector, block and block position indices ($ik$, $ib$, $ip$) respectively. We notice that the observation time is proportional to $t = (ib * N_{spb} + ip) \times (1 / f_{\rm sampling}) + t_0$ and so we can write
\begin{eqnarray}
I_{iwd}(ik,ib,ip) = & \pm C^{ik} P^{ik}_{p(ib)} I^{\rm sky}_{p} ({\rm OPD}_{iwd}(ib,ip)) \\ 
    & + A_{ik}^{\rm atm} \times I^{\rm atm}(ik, ib,ip) \\
    & + A_{ik}^{\rm elec} \times R^{\rm elec}(t) \\ 
    & + ZL^{ik}(ib) + A_{ik}^{\rm M}(ib) \times B_{\rm M}^{ik}(ip) \\
    & + I^{k}_{sl}(ik,ip)
\label{eq:array_treatment_of_interferograms}
\end{eqnarray}
where $iwd$ define either the \fwd\ or \bwd\ raw interferograms.
We define the offset path different as a function of time as ${\rm OPD}(ib,ip)= ZPD(ib,ip) - 2\Delta d (t)/c$ where $ZPD$ denotes the zero path difference. The atmospheric residual signal is assumed to be common between neighbour detectors to a given scaling factor. The electronic noise is assumed to be common for all detectors with a single scaling factor per scan. The parasitic mirror signal is modeled per block and per detector as a constant offset  $ZL$, and a scaled function, $B_{\rm M}$. Finally, we assume the stray light background signal to be constant with time.
\subsubsection{Pre-processing}
\label{sec:preprocessing}
In order to isolate the real astrophysical signal, $I_{p}^{\rm sky}$, from the rest of components, several processing steps are applied separately for \fwd\ and \bwd\ interferograms:
\begin{enumerate}
    \item A linear fit is subtracted to each interferogram within each block, in order to account for the mirror parasitic contribution. 
    \item The ZPD is computed from the mean interferogram for each detector. 
    \item The mean interferogram per detector and per scan is subtracted to all (\fwd\ or \bwd) interferograms. In this way, the stray light contribution is subtracted in OPD space. Strong sources are masked before computing the mean interferogram.
\end{enumerate}
\subsubsection{Correlated noise removal}
\label{sec:corr_decorr}
The atmospheric and other correlated noise components can be subtracted by applying a principal component analysis (PCA) decorrelation algorithm. This analysis is run by comparing the data from all the different detectors ($ik$) in a given array, and deriving common PCA templates accounting for both detector and electronic noise components. In practice we consider the first 5 PCA components, which seem to account for most of the unwanted components in the data. \\ \\
We show in Fig.~\ref{fig:kiss_decorr} an example of the full processing on the KISS interferograms, taken from a Jupiter scan. We show the \fwd\ raw interferograms for one detector and block of data (cyan) after correction of the mirror parasitic signal. In blue we show the mean interferogram, which accounts mainly for the stray light background signal. In green, we show the residual atmospheric and electronic noise, which show an oscillatory behaviour. The PCA decorrelated interferogram is shown in red. We observe that there are various order of magnitude in amplitude with respect to the raw data.
\subsubsection{Projection of interferograms on the sky and spectral inversion}
The processed interferograms for all detectors in one array are jointly projected into the sky in OPD space using a simple co-addition method to obtain interferometric cubes $\hat{I}_{\pm}^{\rm  sky, iwd}(\theta,\phi, {\rm OPD})$ for the \fwd\ and \bwd\ interferograms. A total of 801 bins in OPD were selected to ensure a good representation of the interferograms and in particular in the region of the maximum. This also allows us to fully cover the frequency range of interest for KISS. 
\\
Combining the \fwd\ and \bwd\ interferogram cubes and taking the Fourier transform we can obtain an estimate of the spectrum of the sky:
\begin{eqnarray}
    C H(\nu) &T^{atm}(\nu) \ {\hat{S}}^{\rm sky}(\theta, \phi, \nu) = \\
    &TF^{-1}\left[ \frac{1}{2} \left({\hat{I}}^{\rm sky, fwd}(\theta,\phi, {\rm OPD})+ {\hat{I}}^{\rm sky, bwd}(\theta,\phi, {\rm OPD}) \right) \right],
\end{eqnarray}
where $C$ defines the overall calibration coefficient after the KIDs combination.
\subsection{Calibration}
\label{sec:calspectro}
As for the continuum case, the spectroscopic calibration was made using Jupiter scans and assuming the Jupiter SED model from Section \ref{sec:absphcal}.  As we have only considered compact sources we have decided not to rely on bandpass reconstruction and as so we define a combined spectral calibration coefficient, $C (\nu) = C H(\nu)$. These coefficients per frequency are computed from the flux of Jupiter at each frequency as measured on the spectral cubes defined above. In practice, we construct spectral cubes for the two arrays for each of the scans of Jupiter. We compute the flux of Jupiter at each frequency by fitting the frequency map to a 2D Gaussian. The obtained flux is corrected for atmospheric transmission using an ATM model \citep{2019adw..confE..36P} and the on-site measurements \citep{OTPWV_measurements} of precipitable water vapour (PWV). These PWV measurements are taken by a GPS antenna located at $\sim$1\,km of the telescope, which obtains 4 rapid measurements and 1 precise measurement per hour. Finally, after comparing the measured flux with the expected Jupiter SED we obtain the spectral calibration coefficients.\footnote{In principle, the ATM model could be used together with a skydip observation in order to provide an independent calibration estimate. However, we have observed different near (atmosphere) and far (targets) field responses. Therefore, we decided to calibrate using only the planets.}
\\ \\
These coefficients as well as the best-fit FWHM are shown in Fig.~\ref{fig:spabscal}. In the top plots we show both individual scans (grey lines) as well as mean (blue) and median (red) spectral calibration coefficients. In the bottom plots we show the measured FWHM as a function of frequency. We observe that the FWHM values from spectroscopic data are consistent with the expected behaviour in frequency as derived from continuum measurements (blue line) from Section~\ref{sec:specastro}. The highest frequency point is already too close to the 180\,GHz water vapour line, which implies a much larger and poorly defined calibration factor at that frequency range. We also see great variations on the beam size across scans. Both issues are especially severe for array B, as can be seen in the two panels on the right. The later issue suggests that this point is not only noisier because of the water vapour line, but might also be biased.

\section{Continuum and spectral maps of selected astrophysical targets}
\label{sec:specastro}
We have performed sourced-centered azimuth-elevation raster scans to map 
the sky sources. In practice, we perform two consecutive
subscans and then re-point the telescope to follow the source motion. The
main properties of the scans are the center azimuth and elevation positions,
the map size in the two axes, the step between two consecutive subscans and
the scanning velocity. These observations are prepared and controlled using
a new Graphical User Interface (GUI) customized for KISS. This GUI has been
optimized to be used jointly with the control software of the QUIJOTE telescopes,
based on Twincat \citep{quijote_controlsoftware}. One of the main tasks
of this GUI is to send telescope status and position to the KISS acquisition
system in order to control the tuning procedure of KIDs. A complete
description of this tuning procedure is presented in \cite{bounmy2022}.
\subsection{Moon observations}
We took a total of 284 raster scans of the Moon during the KISS observation 
run. Each of them lasted for 10 minutes, accounting for almost 50 hours at 
the end. The sizes of the maps varied greatly, with most of them being between 
$100^\prime\times100^\prime$ and $200^\prime\times200^\prime$. 
Most of these scans show issues, specially before November 
22$^{\rm nd}$ 2019, when the first complete pointing model version was 
applied\footnote{The pointing model is described in depth in 
Appendix~\ref{appendix:pointing_model}.}. After discarding data 
showing issues we end up with a final set of 52 Moon rasters. 
These Moon rasters were taken on 5 different dates between November 22$^{\rm 
nd}$ 2019 and November 9$^{\rm th}$ 2020.
\subsubsection{Moon photometry}
\label{sec:moon_photometry}
\begin{figure}
    \centering
    \fig{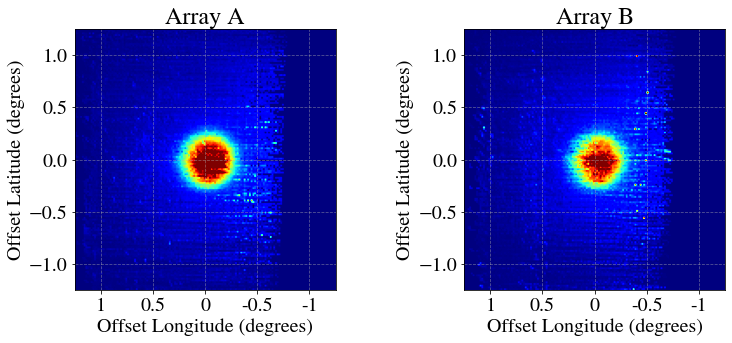}{1\linewidth}{}
    \caption{Continuum maps (left/right: A/B arrays) for a Moon 
    observation taken December 9$^{\rm th}$ 2019. This is the
    brightest observation from our sample. We measured the offset
    on the obtained Moon image with respect to its expected position, 
    in order to look for residual deviations in the pointing model.
    These are only found at the sub-arcminute level, proving that the
    pointing model (which was obtained at the TOD level - see 
    Appendix~\ref{appendix:pointing_model}) works well.}
    \label{fig:moon_example_scan_818}
\end{figure}

\begin{figure}
    \centering
    \fig{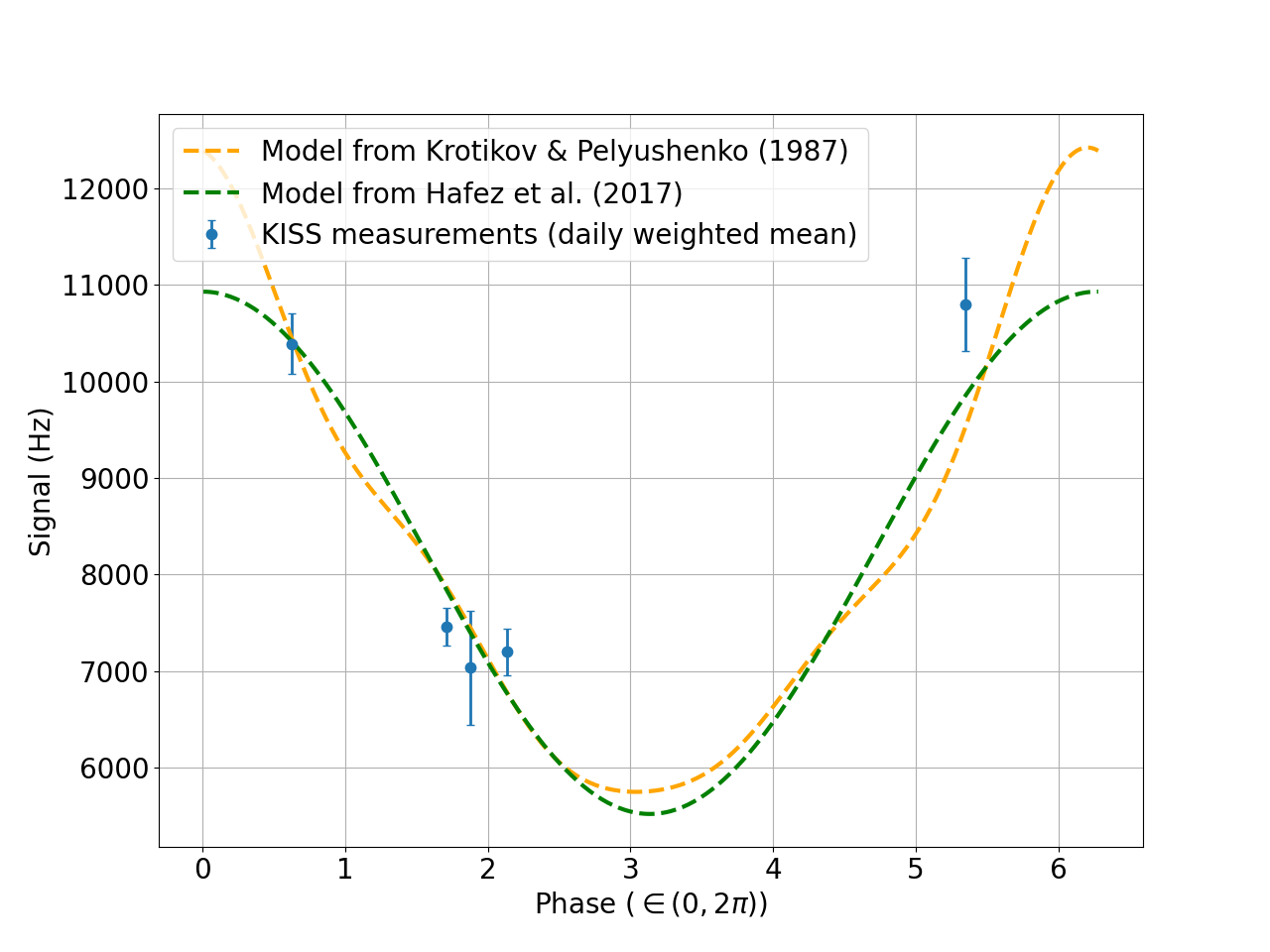}{1\linewidth}{}
    \caption{Comparison between the brightness temperature obtained for the Moon from the data and the models from \cite{old_moon_model} and \cite{hafez_moon_model}. The models have been scaled from brightness temperature to frequency units by a constant factor, plus an offset value. This has been computed as the one minimizing the quadratic sum of the differences w.r.t. the data along the full phase. The five different observation dates account for a reduced coverage of the phase. Data for KIDS array A.}
    \label{fig:moon_phase}
\end{figure}

In Fig.~\ref{fig:moon_example_scan_818} an example photometry map from one of the 
Moon rasters is shown. This scan corresponds to the brightest Moon observation within
our sample, while the Moon was in its full phase. Because of that, the Moon
is evenly illuminated by the Sun and its brightness can be assumed as uniform
within its disk. That implies that, when studying the angular size of the Moon 
emission, a Gaussian fit in both directions is enough 
when the Moon is in its full phase. Thus, we find the FWHM of the best 
Gaussian fit to be consistent to the expected Moon angular size (30 arcminutes). When 
using the FWHM values obtained for the both two axes ($a$, $b$), we find an 
ellipticity value of just  $e=1-b/a=0.024$.
The Moon flux is estimated using aperture photometry. The median signal within a ring with radii 20$\sqrt{2}$ and 40 arcminutes, which accounts for the background signal, is subtracted to the mean signal within a disk of radius 20 arcminutes. The standard deviation of the signal within the former region is used to estimate the uncertainty of the measurement. We then studied these flux values as a function of time, and compared them to the models for the lunar phase from~\cite{old_moon_model} (hereafter, \citetalias{old_moon_model}) and \cite{hafez_moon_model} (hereafter, \citetalias{hafez_moon_model}). For this, we assumed an effective frequency of 150\,GHz for the Moon observations, close to the central frequency of KISS:
\begin{eqnarray}
    T[{\rm K}] &= 227 + \sum_{i} T_i\cos(i\times\omega t - \phi_i);\qquad{\rm from\quad KP14} \\
    T[{\rm K}] &= 208 - 80 \times \cos(\omega t - 14^\circ);\qquad{\rm from\quad H14}
\end{eqnarray}
where $\omega$ is the angular frequency of the Moon cycle (360$^\circ$ / 29.35\,days = 12.26$^\circ$\,days$^{-1}$). $t$ is the elapsed time since a reference New Moon, which we selected as 05:31:31.2 from 12$^{\rm th}$ December 2019. The parameters defining the \citetalias{old_moon_model} model are $i\in(1,2,3,4)$, $T_i\in(107, 19, 15, 7)$\,K and $\phi_i\in(14, 26, 20, 34)$\,degrees. The model is scaled by a constant value to obtain the calibration factor between brightness temperature and the signal in frequency units. An offset constant value is also taken into account. In Fig.~\ref{fig:moon_phase} we show the fitted data and best-fit models. The two calibration factors for KIDS array A 
are $27.5\pm6.3$\,Hz\,K$^{-1}$ for the \citetalias{old_moon_model} model and $33.8\pm6.6$\,Hz\,K$^{-1}$ for the \citetalias{hafez_moon_model} model, the two being consistent within $1\sigma$. The same factors for KIDS array B are $24.1\pm4.18$\,Hz\,K$^{-1}$ and $29.4\pm4.4$\,Hz\,K$^{-1}$, respectively, again consistent within $1\sigma$. Considering that the laboratory measurements of the gain returned values at the 1000\,Hz\,K$^{-1}$ range, then the deficit is around a factor 30-40. 
\begin{figure*}
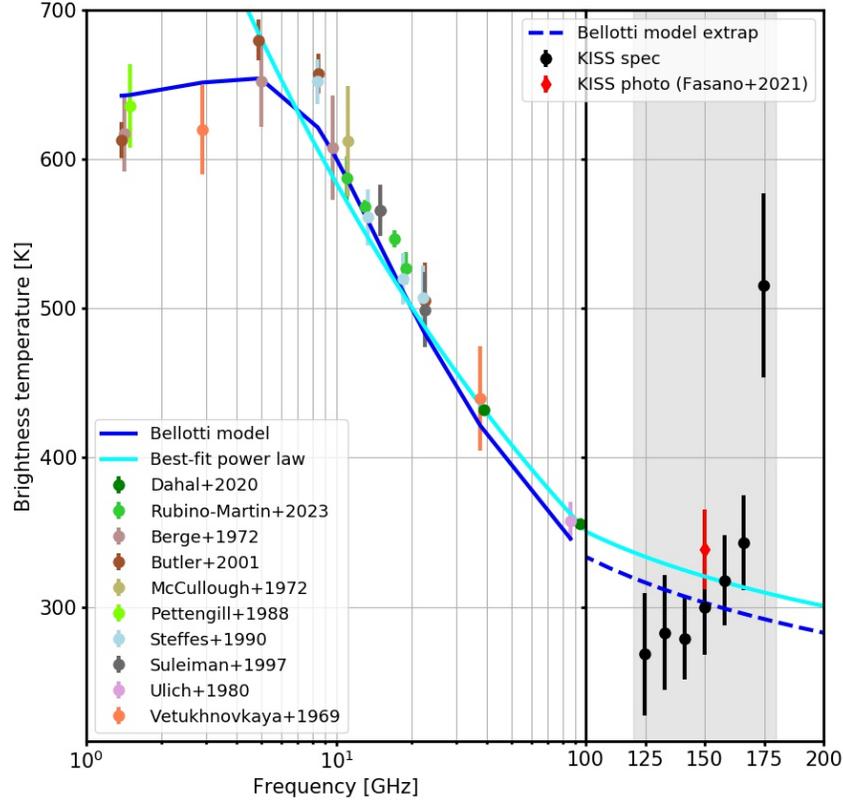

\centering
\fig{figures/Venus_spectrum_KISS.jpg}{0.7\linewidth}{}
\caption{Comparison between the Venus continuum (light red dot) and spectrum (black dots) obtained from KISS  with ancillary data and models. At the edges of the KISS band we expect systematic uncertainties to be large as the transmission is low. This explains why the KISS highest frequency point is significantly above the model. A detailed description of the former is given in \cite{fasano-calib}. We also include recent results from the QUIJOTE experiment in the 10 to 20~GHz frequency range~\citep{2023MNRAS.519.3383R}. \label{fig:venuspectrum}}
\end{figure*}
These values are roughly consistent with the one from Jupiter obtained in Section~\ref{sec:absphcal} (Fig.~\ref{fig:phabscal}), $35.7\pm2.6$\,Hz\,K$^{-1}$, the worst of them being 2.4$\sigma$ away. The agreement is better with array A than B. The lower gain could be pointing to a deficit of signal in the KIDs response because of the extended nature of the Moon, compared to that from Jupiter, or to non-linearities in the instrument response when under large signals. It can be also due to an over-subtraction of the signal when calculating the mean over the full FoV to estimate the baseline level, because of the larger Moon size. These results confirm the issue with the optical coupling between the QUIJOTE telescope and the KISS instrument.
\\ \\
KISS spectral maps and data from the Moon were already presented in \cite{fasano-moonspectra}. In that case, a notch filter was included to discard the emission from the 180\,GHz water vapour line, while being able to observe from 200 to 300\,GHz. The observations described in this work, on the other hand, completely filtered out all emission above 180\,GHz. Below this frequency, both spectral analyses are consistent.
\subsection{Venus spectra}
We obtained 40 Venus scans after the final pointing model was in place. These account for almost 7 hours of observations, each scan being around 10\,min long and covering a $90^\prime\times90^\prime$ region. Continuum results using KISS for Venus were published already in \cite{fasano-calib} and are presented here for the record in Figure~\ref{fig:venuspectrum} in bright red. This is compared with several ancillary observations at radio frequencies between 1 and 100\,GHz, showing a consistent behavior. We also show for comparison (as in \citealt{fasano-calib}) the \cite{bellotti_venusmodel} model (blue solid and dashed line) and simple best-fit power law (cyan data). Notice that we also included in the figure data from recent measurements with the QUIJOTE telescope in the 10 to 19~GHz range \citep{2023MNRAS.519.3383R}.\\
\\
We have obtained the spectrum from the same Venus data using the approach described in Section~\ref{sec:spectroscopy}. The obtained spectral density is shown as black dots in the figure. Uncertainties were computed from the dispersion between scans and also include the absolute calibration uncertainties presented in Sect.~\ref{sec:calspectro}. The uncertainties are dominated by the latter and in particular at the edges of the KISS band. This is particularly obvious at high frequency for which systematic uncertainties dominate the signal. Overall our measurements are consistent with the two aforementioned models proposed at the 95 \% C.L. C
\section{Conclusions and Perspectives}
\label{sec:conclu}
This is the last of several papers describing the performance of the KISS instrument \citep{fasano-moonspectra, fasano-nika2, fasano-ltd, fasano-calib}. We have shown that KIDs and interferometers (and more specifically, a MPI) can be used together to provide simultaneous photometric and spectral measurements of astrophysical targets in the millimetre (between 100 and 180\,GHz). The beam properties of the instrument matched those from its design phase, achieving FWHM values of 7--8$^\prime$ at 150\,GHz over the large KISS FoV (1\degr). The recovered focal plane geometry also followed the one measured at the laboratory, with $\sim$80\% of the KIDs working normally.
\\ \\
The main issue that we encountered during the observation phase of the instrument was a deficit in its response with respect to the one expected. This deficit is of the order of a factor 30, and is expected to be due to a miscoupling between the instrument and the telescope. 
It is possible that despite the fact that the QUIJOTE telescope was designed up to mm wavelength aging and environmental conditions have degraded its performance. Furthermore, we notice, first, that such an issue was not observed during KISS laboratory testing, and second, that no problem was found during CONCERTO commissioning phase \citep{2024arXiv240615572H}. As so we are confident that the deficit of response is not related to the instrumental design of KISS.
\\ \\
This degraded response prevented KISS from fulfilling its second main objective: to provide measurements of the SZe from large GCs. However, it did not prevent KISS from being a reliable test-bench for CONCERTO \citep{concerto}. This was particularly the case for the optical configuration and the readout, the raw calibration methodology and the development of the data analysis pipeline.
\\ \\
Jupiter and Venus were observed with sufficient signal-to-noise ratio as to measure their SED. 
Moon observations were also used to show the slight differences encountered when observing an extended source. The phase of the Moon is well recovered from our data, showing that the calibration factor did not vary greatly during our observing run. Finally, our results show the promising prospects of the combination of KIDs with interferometers to provide simultaneous photometric and spectral measurements.

\begin{acknowledgments}
We thank the staff of the Teide Observatory for invaluable assistance in the commissioning and operation of QUIJOTE.
The {\it QUIJOTE} experiment is being developed by the Instituto de Astrofisica de Canarias (IAC),
the Instituto de Fisica de Cantabria (IFCA), and the Universities of Cantabria, Manchester and Cambridge.
Partial finantial support was provided by the Spanish Ministry of Science and
Innovation under the projects AYA2017-84185-P and PID2020-120514GB-I00.
We thank Angelo Mattei of the Mechanical Workshop Service of the INFN Rome 1 Section for manufacturing the HDPE lenses. MFT acknowledges funding from the French Programme d’investissements 
d’avenir through the Enigmass Labex, and support from the 
Agencia Estatal de Investigación (AEI) of the Ministerio de 
Ciencia, Innovación y Universidades (MCIU) and the European 
Social Fund (ESF) under grant with reference PRE-C-2018-0067.
This project has been partially supported by the European Research Council (ERC) under the European Union’s Horizon 2020 research and innovation programme (grant agreement No 788212) and by the LabEx FOCUS ANR-11-LABX-0013.
\end{acknowledgments}
%
\software{astropy \citep{astropy:2013, astropy:2018, astropy:2022},  matplotlib \citep{Hunter:2007}, numpy \citep{harris2020array}, scipy \citep{2020SciPy-NMeth}}

\appendix
\section{Interferometric equation derivation}
\label{appendix:interferometric_derivation}
Assuming a monochromatic input signal and focusing on $S_{1}$ and following the notations of Fig.~\ref{fig:kissmpiconcept} we can write the electric field for each of the previously described steps:
\begin{eqnarray*}
&\vec{E}_{0}  =  \frac{\vec{E}_{S_{1}}}{\sqrt{2}} \cos{ \lbrace 2 \pi \nu_{0} \left(t - \frac{x}{c} \right) \rbrace } \left( 0, 1, 0 \right)  \\
&\vec{E}_{+1}  =  \frac{\vec{E}_{S_{1}}}{2\sqrt{2}} \cos{ \lbrace 2 \pi \nu_{0} \left(t - \frac{x}{c} \right) \rbrace }  \left( 0, 1, 1 \right)  \\
&\vec{E}_{+2}  =  \frac{\vec{E}_{S_{1}}}{\sqrt{2}} \cos{ \lbrace 2 \pi \nu_{0} \left(t - \frac{x}{c} \right) \rbrace } \left( 0, 1, -1 \right) 
\end{eqnarray*}
Reflections on $M_{1}$ and $M_{2}$ mirrors lead to $\vec{E}_{-1} = - \vec{E}_{+1}$ and $\vec{E}_{-2} = - \vec{E}_{+2}$. For the combined beam we have $\vec{E}_{3} = \vec{E}_{-1}+ \vec{E}_{-2}$ and therefore 
$$ \langle |\vec{E}_{3}|^{2} \rangle = \langle |\vec{E}_{1} + \vec{E}_{2}|^{2} \rangle =  \langle |\vec{E}_{1}|^{2} \rangle + \langle |\vec{E}_{2}|^{2} \rangle + 2 \langle |\vec{E}_{1}|  |\vec{E}_{2}| \rangle.$$ 
After the output polariser, $P_{2}$, the observed intensity in the KIDs' arrays can be written as $I+ = 2  |\vec{E}_{3}.\vec{e_{y}}|^{2}$ and  $I+ = 2  |\vec{E}_{3}.\vec{e_{x}}|^{2}$, where $\vec{e_{y}}$ and $\vec{e_{x}}$ are unitary vectors along the $x$ and $y$ axis. Therefore, we can write 
$$I_{\pm} = \frac{P_{S_{1}}}{4} \left[ 1 \pm \cos{ \lbrace 2 \pi \nu_{0} \frac{2 \Delta d (t)}{c} \rbrace } \right],$$ 
where $2 \Delta d (t)$ is the optical path difference between the two roof mirrors at time, $t$. \\
\section{Laboratory measurements}
\label{appendix:lab_measurements}
\begin{figure*}[h]
	\centering
	\fig{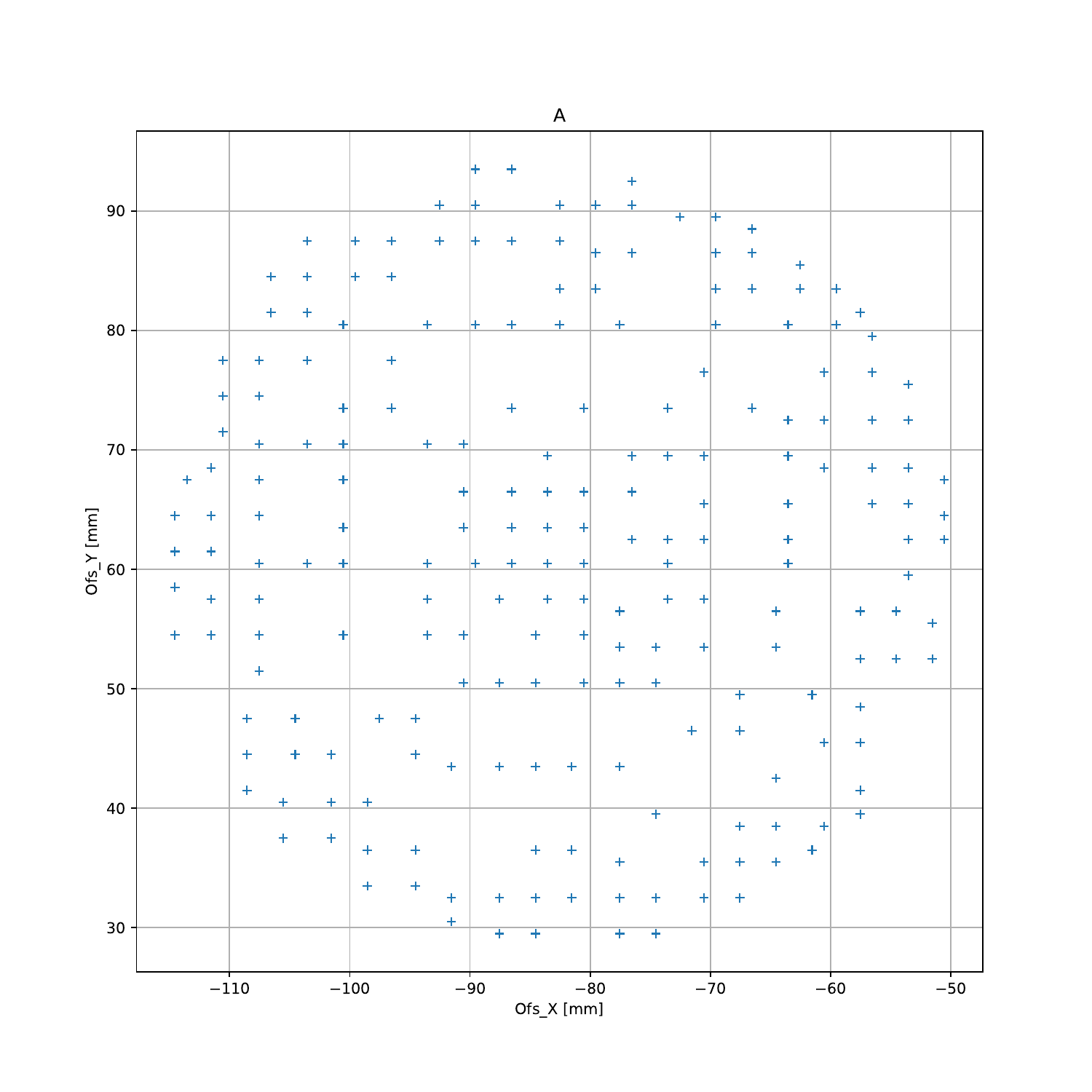}{.45\textwidth}{}
	\fig{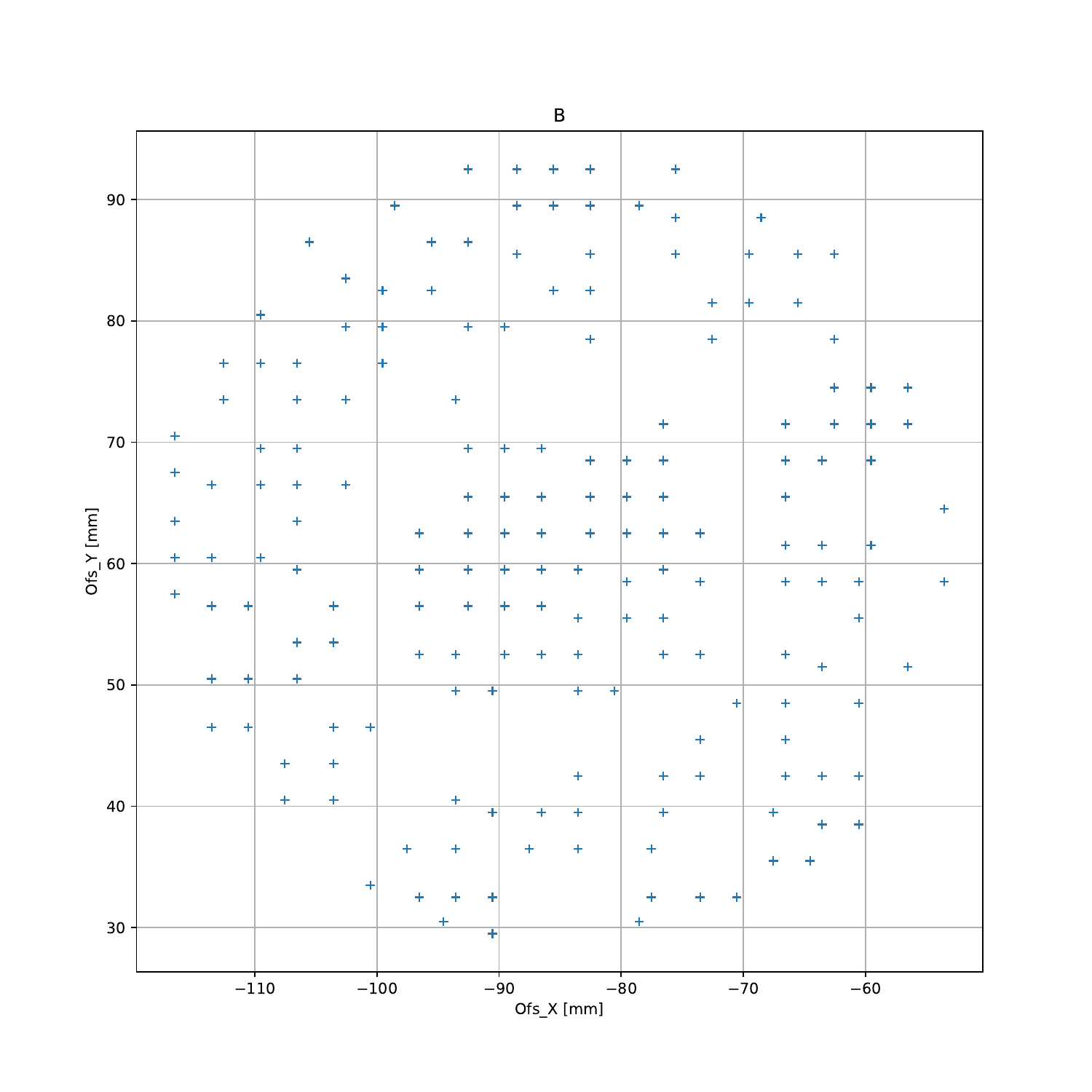}{.45\textwidth}{}
	\caption{Geometry of the focal plane, from laboratory measurements. KID offset position for array A (left) and B (right).}
	\label{fig:labfocalplane}
\end{figure*}

\begin{figure*}[h]
	\centering
	\fig{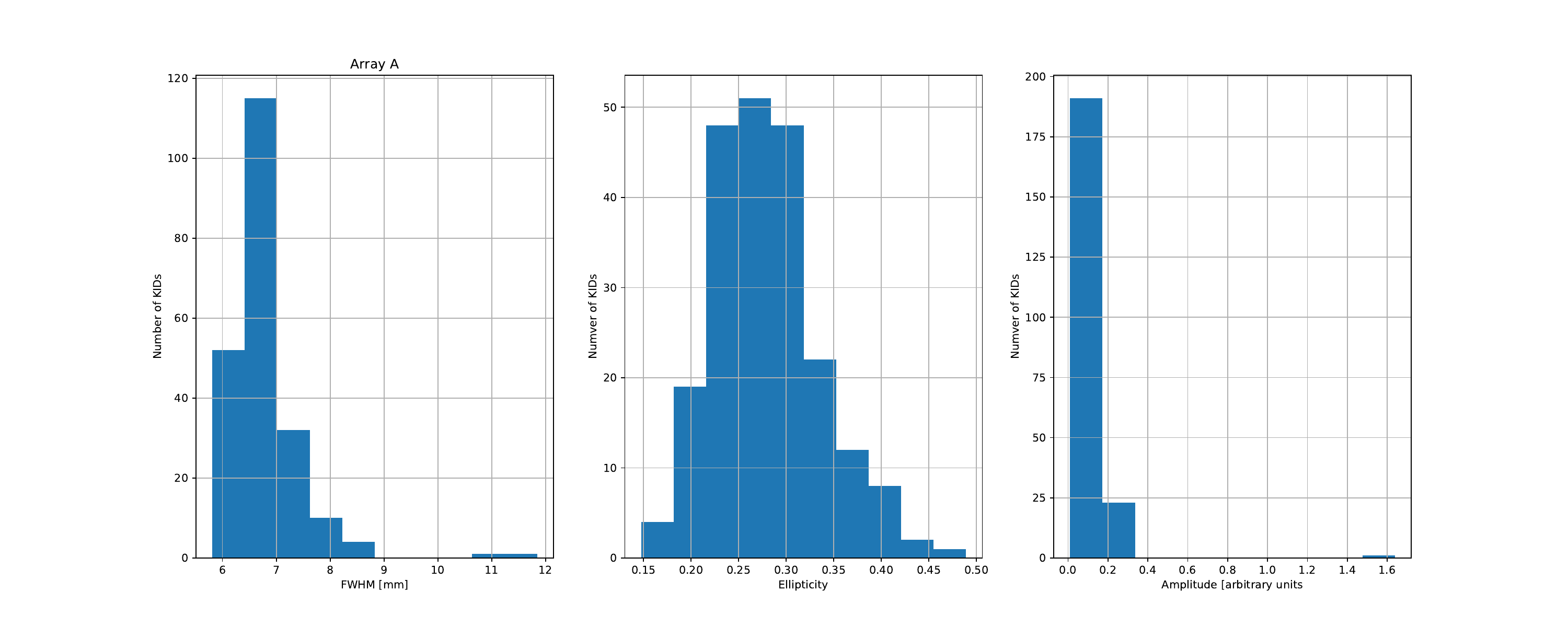}{1\textwidth}{}
	\fig{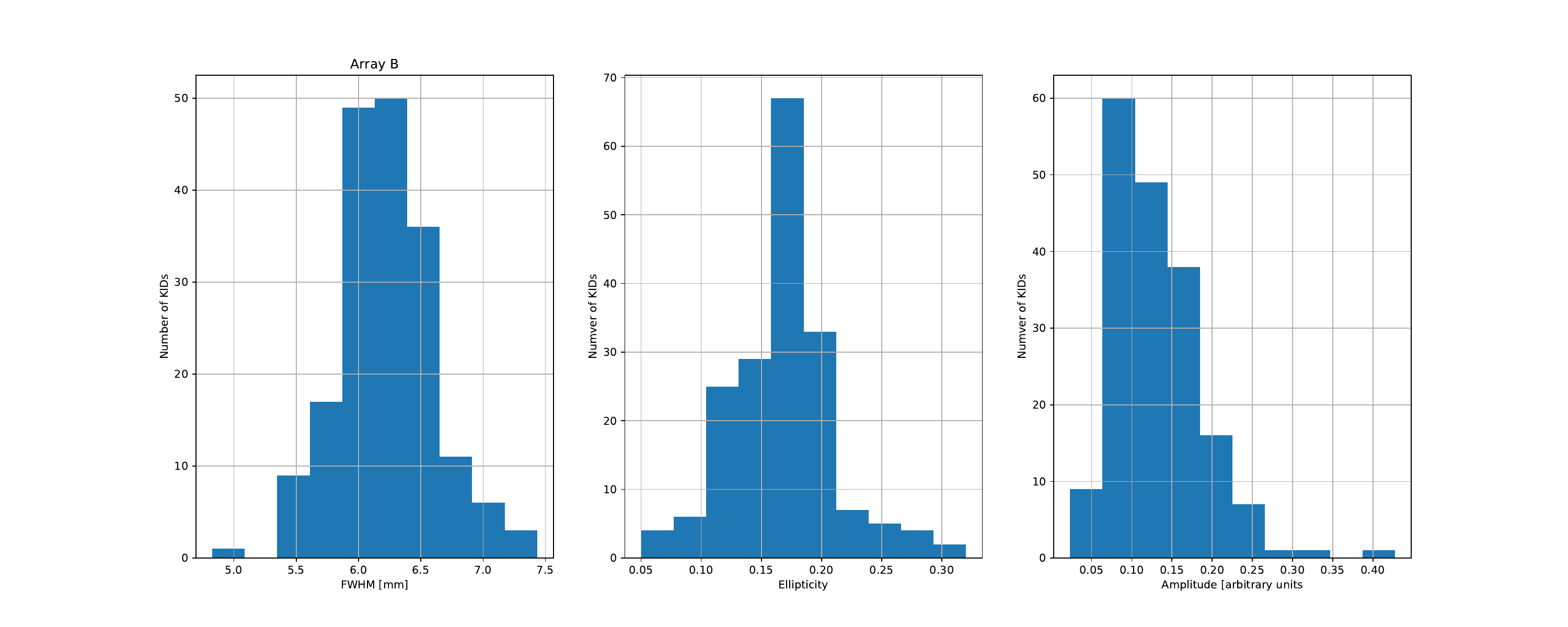}{1\textwidth}{}
	\caption{Measurements of KIDs FWHM, ellipticity and response / amplitude (from left to right) obtained from laboratory measurements. Results for arrays A (top) and B (bottom).}
	\label{fig:labkidproperties}
\end{figure*}

During laboratory tests, we were able to reconstruct the position of the KIDs in the focal plane using a background sky simulator coupled to a fake planet. The sky simulator, consisting of cold head coupled to a pulse tube, produces a nearly homogenous  sky temperature. The fake planet, made of a metal ball attached to a metal arm by a kevlar wire, can be displaced vertically and horizontally to mimic raster scans, which were used to derive the focal plane geometry. For each scan the raw data were projected into maps per KID and these were fitted to a 2D Gaussian to derive the KID position in the focal plane, the beam FWHM and ellipticity and the relative KID response. Fig.~\ref{fig:labfocalplane} shows the reconstructed position of the KIDs in the focal plane for array A (left) and B (left). The 1 degree FoV is fully sampled, with better coverage in the center part and on the outer region. In Fig.~\ref{fig:labkidproperties} we present the main measured properties of KIDs in terms of beam FWHM and ellipticity, and amplitude. We observe that the KIDS have similar properties with nominal behaviour, about 6-7 mm FWHM. The small excess of ellipticity is mainly due signal from the wire.
\section{Pointing model}
\label{appendix:pointing_model}
\begin{figure}
    \centering
    \fig{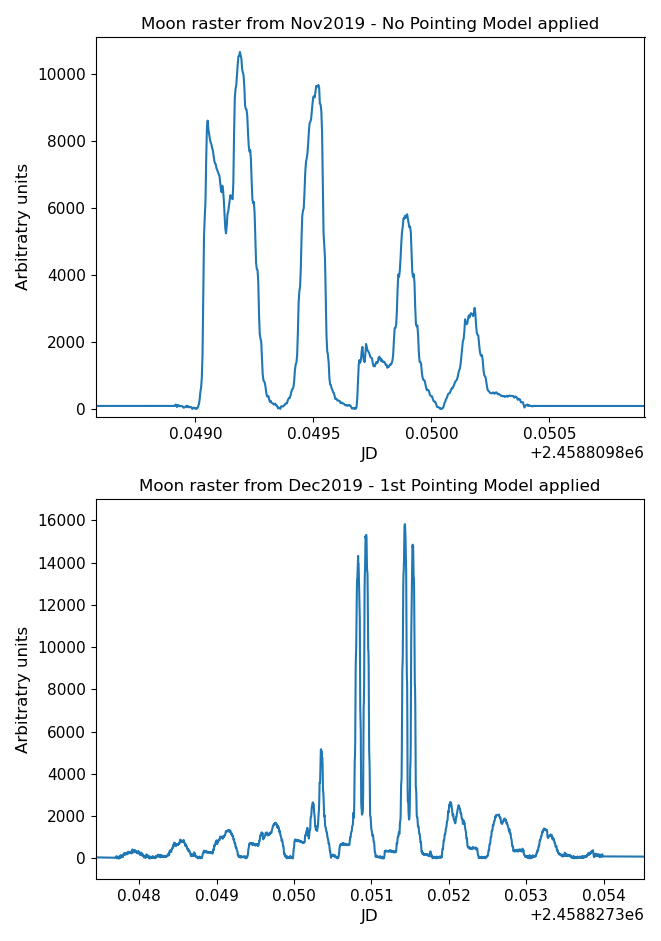}{0.6\linewidth}{}
    \caption{Top (bottom): example of a Moon raster before (after) applying the first version of the pointing model. This version only used Moon data, and fixed the azimuth and elevation offsets between the sky and the telescope encoder. The improvement is already noticeable, as the expected double-peaked behavior emerges after applying this preliminary version of the pointing model. The angular, steep features close to the peaks, typical of an imperfect pointing model, also lose prominence after applying this first version of the pointing model.}
    \label{fig:PM_before_after_Moon}
\end{figure}

\begin{figure*}
    \centering
    \fig{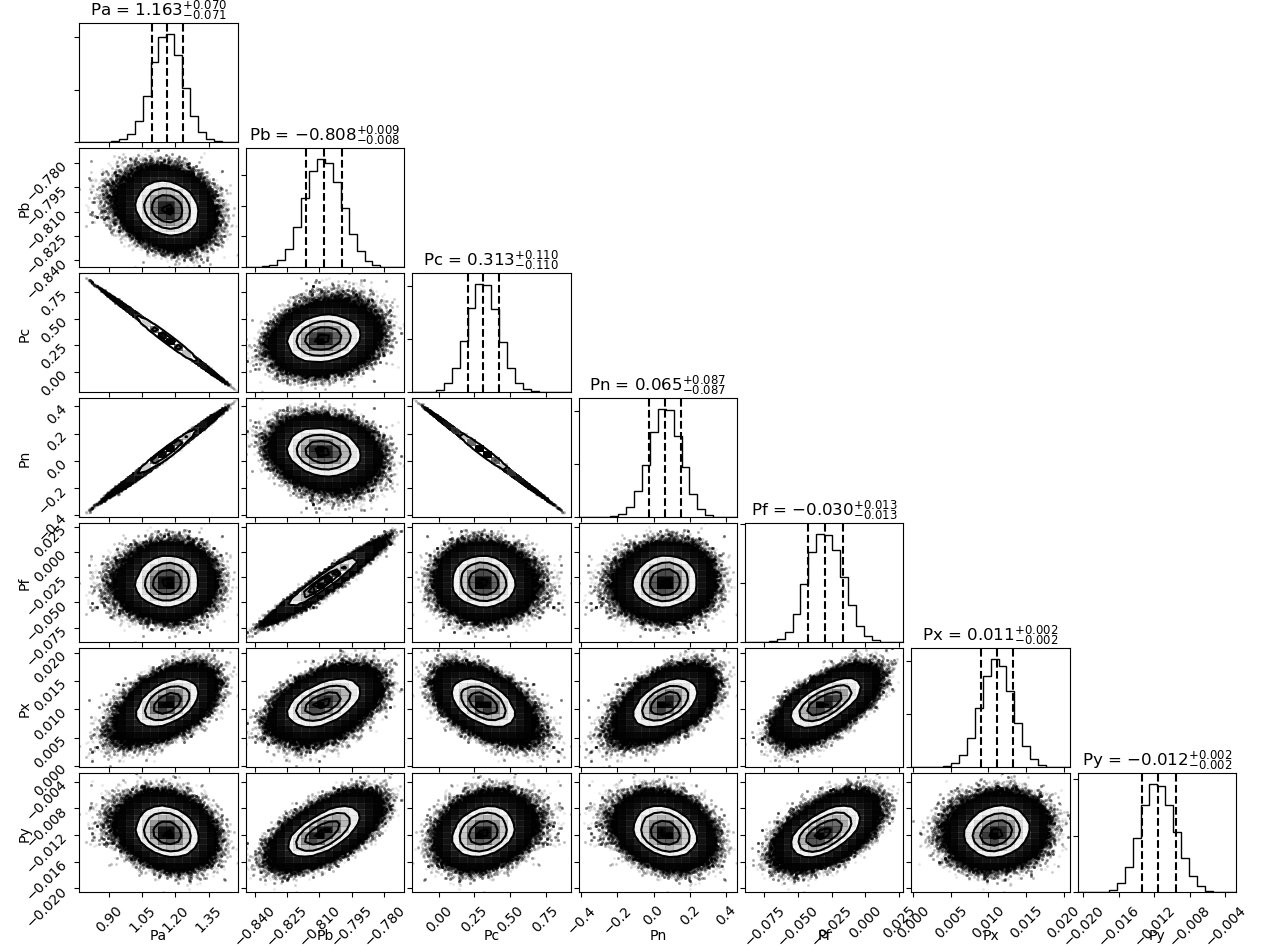}{1\linewidth}{}
    \caption{Corner plot showing the posteriors for the final values of the pointing model, using data from Moon, Jupiter and Venus observations. There are degeneracies between the azimuth-related ($P_a$, $P_c$, $P_n$) and elevation-related ($P_f$, $P_b$) parameters due to the lack of calibrators for every azimuth-elevation position. Units are in degrees.}
    \label{fig:PM_cornerplot_7params}
\end{figure*}

The pointing was reconstructed using measurements of the Moon, Jupiter and Venus. A first pointing model was obtained from Moon observations only (Fig.~\ref{fig:PM_before_after_Moon}) and then it was refined using the observations of the planets. The final pointing model consists of 7 parameters (Fig.~\ref{fig:PM_cornerplot_7params}). The description of these parameters can be found in \cite{wallace2008}. The pointing model parameters were derived from the time ordered data of detector KA005, which is at the center of array A. The astrophysical signal was modeled assuming a 2D Gaussian beam with FWHM equal to 30 arcminutes for the Moon and 7 arcminutes for the planets. 
\bibliographystyle{aasjournal}
\bibliography{bibliography}{}
\end{document}